\documentclass[a4paper,fleqn,usenatbib,useAMS]{mnras}

\usepackage{graphicx}	
\usepackage{amsmath}	
\usepackage{amssymb}	
\usepackage{multicol}        
\usepackage{multirow}
\usepackage{bm}		
\usepackage{pdflscape}	

\usepackage[T1]{fontenc}
\usepackage{ae,aecompl}
\usepackage{mathptmx}
\usepackage{longtable}
\usepackage{array}
\usepackage{booktabs}

\title[Evaluating Schwarzschild modelling with Illustris]{Evaluating the ability of triaxial Schwarzschild modelling to estimate properties of galaxies from the Illustris simulation}

\author[Y. Jin et al.]
{Yunpeng Jin$^{1,2}$\thanks{E-mail:ypjin@nao.cas.cn}, Ling Zhu$^{3}$\thanks{E-mail:lzhu@shao.ac.cn}, R. J. Long$^{4,1,5}$, Shude Mao$^{4,1}$, Dandan Xu$^{4}$, Hongyu Li$^{1}$, \and Glenn van de Ven$^{6,7}$\\
$^{1}$National Astronomical Observatories, Chinese Academy of Sciences, 20A Datun Road, Chaoyang District, Beijing 100101, China\\
$^{2}$University of Chinese Academy of Sciences, Beijing 100049, China\\
$^{3}$Shanghai Astronomical Observatory, Chinese Academy of Sciences, 80 Nandan Road, Shanghai 200030, China\\
$^{4}$Department of Astronomy, Tsinghua University, Beijing 100084, China\\
$^{5}$Jodrell Bank Centre for Astrophysics, School of Physics and Astronomy, The University of Manchester, Oxford Road, Manchester M13 9PL, UK\\
$^{6}$Department of Astrophysics, University of Vienna, T\"urkenschanzstrasse 17, 1180 Vienna, Austria\\
$^{7}$European Southern Observatory, Karl-Schwarzschild-Str. 2, 85748 Garching b. M\"unchen, Germany
}

\date{Accepted XXX. Received YYY; in original form ZZZ}
\pubyear{2018}

\begin{document}
\label{firstpage}
\pagerange{\pageref{firstpage}--\pageref{lastpage}}
\maketitle

\begin{abstract}
We evaluate the capabilities of Schwarzschild's orbit-superposition method by applying it to galaxies from the large scale, high resolution Illustris simulation. Nine early-type galaxies with a range of triaxiality are selected, and we create mock integral field unit data for five line-of-sight projections of each galaxy. Each of the 45 mock data sets is taken as an independent observed galaxy. Using van den Bosch's 2008 triaxial Schwarzschild implementation, we assess model estimates of various galaxy properties, covering mass profiles, intrinsic shapes, stellar orbit distributions and velocity anisotropies. Total mass within $\overline{R_{\rm e}}$ is recovered well with average deviations within $\pm15$ percent. Stellar mass is underestimated by $\sim24$ percent and dark matter overestimated by $\sim38$ percent (assuming an NFW dark matter profile and allowing for degeneracy between stellar mass and dark matter mass). Using a gNFW profile, these values improve to $\sim13$ percent for stellar mass and $\sim18$ percent for dark matter. Axis ratio estimates show a moderate bias of $\Delta (b/a)=0.07$ and $\Delta (c/a)=0.14$ ($a\ge b\ge c$). Distributions of the orbit circularities $\lambda_z$ and $\lambda_x$, representing rotation about the minor and major axes, are well reconstructed. Separating orbits into thermal categories, our models match the average fractions of these categories to within $10$ percent. Velocity anisotropy is well estimated with values matching in the inner regions but becoming slightly radially biased in the outer regions. Overall, the galaxy property estimates we obtained using Schwarzschild modelling are not implausible and are representative of the simulated galaxies we modelled.\\

\end{abstract}



\section{Introduction}
\label{sec1}
The challenge we face in galactic astronomy is how to analyse observations of a galaxy and not only understand what the observations themselves are telling us but what can also be understood or implied about the galaxy's properties that are not directly observable. Such properties include for example the dark matter distribution, orbit structures, variations in velocity dispersion anisotropy and so on.

In general (our own Galaxy being a major exception), we work with surface photometry and integrated line-of-sight spectroscopic data as provided by an Integral Field Unit (IFU). In recent years, good quality data from IFU surveys such as SAURON \citep{Bacon2001}, $\rm ATLAS^{3D}$ \citep{Cappellari2011}, CALIFA \citep{Sanchez2012}, SAMI \citep{Bryant2015}, and MaNGA \citep{Bundy2015} has become available. This data is characterised by a two dimensional on-sky spatial position with line of sight kinematics or line strength data. Not only do we wish to understand galaxy properties which are not directly observable but we also have to contend with degeneracies generated by projecting or deprojecting our observed data.

Various modelling schemes have been devised to assist investigations such as ours. The Schwarzschild orbit-superposition technique \citep{Schwarzschild1979} we use in this paper is one of them. Other schemes include solutions of the Jeans equations like the Jeans-Anisotropic-MGE (JAM) models \citep{Cappellari2008}, the particle based Made-to-Measure (M2M) method \citep{Syer1996}, and action-angle distribution functions (e.g.,\citealp{Binney2010}) which have mainly been applied to observations of the Milky Way. All these schemes assume that if a model is able to reproduce the observed data from a galaxy then the model can also be used to determine underlying properties of the galaxy.

JAM is computationally quick to run but, having strong assumptions on galaxy properties, may not produce physically realisable solutions, and may not be able to determine the properties we are interested in. Schwarzschild's method and M2M are more computationally time-consuming but have the advantage that the orbit properties of a galaxy may be assessed and this has contributed to our decision to use Schwarzschild's method. However the solutions obtained from both these methods are heavily dependent on how well the initial conditions represent the galaxy being modelled.

Ideally before using the modelling schemes with real galaxies, the systematics of the schemes for their intended use should be well understood. This concerns not only whether or not observed data are reproduced but also the accuracy to which model estimates of underlying properties can be made. This will be a key feature of our investigation. Note that in order to assess accuracy we will need test galaxies with known properties. We will use mock (test) galaxies taken from a cosmological simulation for this purpose.

It is not unusual for galaxies to be modelled under various simplifying assumptions. For example, dark matter may be ignored, spherical symmetry may be assumed, and galaxy features ignored (e.g. bars and spiral arms). We will address this in part by including dark matter in our models, and by seeking to model early-type triaxial galaxies. Bars and spiral arms in late-type galaxies add modelling complexity through their lack of dynamical equilibrium and are excluded. Axisymmetric models rather than triaxial ones are often constructed to analyse galaxies. The main reason for this is that axisymmetric models have fewer parameters, and so they are simpler and faster to execute. However, in real observations such as MaNGA, many early-type galaxies have twisted velocity fields which can only be created by triaxial galaxies. Recent observations appear to suggest this \citep{Li2018}. In addition, cosmological simulations such as EAGLE \citep{Schaye2015,Crain2015} and Illustris \citep{Vogelsberger2014a,Vogelsberger2014b,Genel2014,Nelson2015} show that many galaxies are in fact triaxial with some being close to oblate or alternatively close to prolate \citep{Velliscig2015,Li2016}.

Previous attempts to assess the systematic behaviour of Schwarzschild's method in estimating specific properties of galaxies include \citet{vdV2008} (internal orbit sructure) and \citet{RvdB2009} (intrinsic shape) who used mock galaxies based on theoretical Abel models \citep{Dejonghe1991,Mathieu1999}. \citet{Thomas2007} considered recovery of mass distributions while \citet{Kowalczyk2017} attempted to recover both mass distribution and orbit anisotropy in mock dwarf galaxies. \citet{Zhu2018b} investigated the recovery of orbit circularity. For other modelling schemes, \citet{Li2016} investigated the behaviour of JAM, and \citet{Long2012} extended the earlier Schwarzschild and Jeans equation work of \citet{Cappellari2006} by providing a M2M mass-to-light comparison.

We now consolidate the statements so far into a set of objectives for our investigation. Using an existing Schwarzschild implementation capable of modelling triaxial galaxies \citep{RvdB2008} together with mock triaxial galaxy data constructed from a cosmological simulation, our objectives are to understand how well Schwarzschild's method is able to estimate the underlying properties of our mock galaxies. We consider four properties in particular,
\begin{enumerate}
\item the mass profile including both stellar and dark matter,
\item galaxy morphology,
\item orbit circularity, and
\item velocity dispersion anisotropy.
\end{enumerate}

To our knowledge, this is the first time a quantitative assessment of the estimation of the four properties has been undertaken for early-type triaxial galaxies using Schwarzschild's method.

The structure of the paper is as follows. In $\S$~\ref{sec2} we explain the approach we will take to address the objectives. In $\S$~\ref{sec3} we introduce Schwarzschild's method while in $\S$~\ref{sec4} we present lower level details covering the stellar and dark matter potentials, intrinsic shape parameters, and parameters used to characterize orbit distributions. Information about the mock galaxies we select is given in $\S$~\ref{sec5}. In $\S$~\ref{sec6}, we present the results of our parameter estimations based on Schwarzschild modelling and compare our model values with the true galaxy values. In $\S$~\ref{sec7}, we discuss any biases revealed in attempting to recover the properties of our mock galaxies. Finally in $\S$~\ref{sec8}, we summarize our research, indicating where we have met our objectives and where outstanding issues remain to be investigated.


\section{Approach}
\label{sec2}
The previous section was concerned with setting out the scientific context and objectives for our investigation. This section describes at a top level the approach we are going to take to address those objectives. To recap, we are using an existing Schwarzschild implementation capable of modelling triaxial galaxies together with mock triaxial galaxy data constructed from a cosmological simulation to try and understand how well Schwarzschild's method is able to estimate underlying properties of our mock galaxies.

We employ the software implementation of Schwarzschild's method described in \citet{RvdB2008}. As indicated in the paper, this implementation draws on design ideas from \citet{Rix1997}, \citet{vdM1998}, \citet{Cretton1999}, \citet{Verolme2002} and \citet{Cappellari2006}. In particular, it is able to model triaxial galaxies using Gauss-Hermite coefficients from line of sight velocity distributions as constraints, and uses orbit dithering to reduce noise in model orbit-based calculations. The software has not been parallelised. Other than a number of small minor changes targetted at improving primarily parameter handling, no changes have been made to the van den Bosch software. The core software implementing the Schwarzschild mechanisms is unaltered.

The gravitational potential for the stellar matter in a triaxial galaxy is straightforward to calculate using the Multi-Gaussian Expansion (MGE) formalism \citep{Emsellem1994a,Cappellari2002}. MGEs are able to handle the multiple viewing angles required by triaxial modelling and their use is an integral feature of the software implementation we have chosen. The shapes of dark matter haloes are hard to determine through observations, so we we use a spherical Navarro-Frenk-White (NFW) profile \citep{Navarro1996}. We treat the central black hole as a point source. The major complication that triaxial galaxies do introduce is that three viewing angles must be taken into account. Axisymmetric galaxies by comparison only have two viewing angles. For the Schwarzschild modelling runs, the viewing angles are treated as free parameters whose values are to be determined.

Our mock galaxies and their observations are taken from the Illustris cosmological simulations \citep{Vogelsberger2014a,Vogelsberger2014b,Genel2014,Nelson2015}. The Illustris project is a series of large-scale hydrodynamical cosmological simulations of galaxy formation, using advanced moving mesh code AREPO \citep{Springel2010} with comprehensive physical process modelling. Star formation is included (e.g., \citealp{Springel2003,Schaye2008}) as well as primordial and metal-line cooling processes (e.g., \citealp{Katz1996,Wiersma2009a}), stellar evolution and chemical enrichment procedures (e.g., \citealp{Wiersma2009b,Few2012}), stellar feedback (e.g., \citealp{Navarro1993,Springel2003}) and AGN feedback (e.g., \citealp{Springel2005,DiMatteo2005}). With these physical process models, Illustris is able to produce galaxies sufficiently realistic for our purposes. We choose nine early type galaxies for our Schwarzschild evaluation. The morphologies are deliberately selected such that we have three galaxies close to being oblate, three close to prolate, and three definitely triaxial. For each galaxy, we construct IFU-like mock brightness and kinematic data at five different sets of viewing angles.

The cosmological parameters adopted in the Illustris simulations and thus implicit in our Schwarzschild modelling are: $\Omega_{\rm m}=0.2726$, $\Omega_{\rm \Lambda}=0.7274$, $\Omega_{\rm b}=0.0456$, $\sigma_8=0.809$, $n_{\rm s}=0.963$, and $H_0=\rm 70.4\ km\cdot s^{-1}\cdot Mpc^{-1}$ \citep{Hinshaw2013}.

On a per galaxy basis, the steps we take are
\begin{enumerate}
\item calculate the true values for the galaxy properties we are interested in from the simulation stellar particle data,
\item construct the mock data, at 5 different sets of viewing angles, to be used as constraints in the Schwarzschild models,
\item for each set of mock data
    \begin{enumerate}
	\item  execute a series of Schwarzschild models varying the free parameters,
    \item  select the best-fitting model matching the constraint data,
	\item  from the best-fitting model, calculate the model values for the galaxy properties,
	\item  compare the model property values with the true property values,
    \end{enumerate}
\item collate the results for the five viewing angle sets and assess what has been achieved.

The method we use for varying the free parameters is an optimised grid search as described in \citet{Zhu2018a}. No amendments to the method or its associated software are required.
\end{enumerate}

\section{The Schwarzschild's method}
\label{sec3}
The Schwarzschild's method is a flexible orbit-superposition method for building dynamical models of galaxies in dynamical equilibrium and was first proposed by \citet{Schwarzschild1979}. Initially, the method was used to create spherical and axisymetric models (e.g., \citealp{Richstone1980,Richstone1982,Richstone1984a,Richstone1984b,Levison1985}). \citet{Schwarzschild1982,Schwarzschild1993} extended the method to more general triaxial shapes. The method has been widely used in dynamical modelling (e.g., \citealp{Merritt1996,Zhao1996,Rix1997,Siopis2000,Hafner2000,vdV2006,Capuzzo2007,RvdB2008,Zhu2018a,Zhu2018b}), and has been applied to more complicated barred galaxies (e.g., \citealp{Wang2012,Wang2013,Vasiliev2015}). The steps we take in modelling a galaxy with Schwarzschild's method are indicated below, and follow the approach in \citet{RvdB2008}.

Creating gravitational potentials for our galaxies is straightforward. Our potentials consist of a stellar potential, a dark matter potential and also a potential generated by a central black hole. For the stellar component, we use the Multi-Gaussian Expansion (MGE) formalism \citep{Cappellari2002}. Within the stellar MGE, we use the triaxial viewing angles to deproject the surface brightness into a three-dimensional luminosity density, and use a constant mass-to-light ratio to turn the density into a mass density from which the potential can be determined.

We create initial conditions for our orbits by sampling from the three integrals of motion defining a triaxial system. Taking the integrals as energy $E$, second integral $I_2$, and third integral $I_3$, these integral triplets ($E, I_2, I_3$) play the same role as ($E, L_z, I_3$) in an axisymmetric system \citep{Binney2008}. In this paper, we take $21\times10\times7$ combinations of ($E, I_2, I_3$) to form initial conditions for our orbits.

Box orbits are crucial for supporting the triaxial shape, so we add further initial conditions specifically for box orbits \citep{Schwarzschild1993,RvdB2008}. As box orbits always touch equipotentials \citep{Schwarzschild1979}, we construct our initial conditions on equipotential surfaces using energy $E$ and two spherical angles ($\theta, \varphi$) giving a further $21\times10\times7$ orbits. Since retrograde stars may exist in elliptical \citep{Bender1988} and lenticular galaxies \citep{Kuijken1996}, we also produce initial conditions which will result in counter-rotating orbits.

In total we have 3 sets of $21\times10\times7$ orbits giving 4410 orbits in our orbit library. The library comprises a typical set of ($E, I_2, I_3$) orbits, a set of counter-rotating orbits ($E, -I_2, I_3$), and the box orbits ($E, \theta, \varphi$).

As in \citet{Cappellari2006} and \citet{RvdB2008}, we dither every orbit to give $5^3$ orbits by perturbing the initial conditions slightly. Once the orbit trajectories have been created the trajectories will be co-added to form a single orbit trajectory in our orbit library. The rationale behind dithering orbits is that the mechanism should reduce the Poisson noise in calculating model observables. In creating the trajectories, we integrate along the orbits in our gravitational potential for 200 orbital periods and aim to achieve a relative accuracy of $10^{-5}$ in energy conservation.

In arriving at the orbit weights, we use surface brightness and luminosity density from a galaxy's MGE together with Gauss-Hermite coefficients \citep{vdM1993,Gerhard1993} of the line-of-sight velocity distribution as constraints. Schwarzschild's method weights orbit contributions to model observables in an attempt to match the observed values. In so doing the model and observed values are divided by the observational error so that a $\chi^2$ comparison is achieved. The weights themselves are determined by the \citet{RvdB2008} implementation using the \citet{Lawson1974} non-negative least squares (NNLS) implementation.

\section{Formulas and Parameters}
\label{sec4}
In this section, we introduce the key formulas and parameters used in our modelling. $\S$~\ref{sec4.1} to $\S$~\ref{sec4.4} deal with the gravitational potential; $\S$~\ref{sec4.5}, orbits; and $\S$~\ref{sec4.6}, error analysis.
\subsection{Stellar potential}
\label{sec4.1}
We construct our stellar potential by using the Multi-Gaussian Expansion (MGE) formalism, the starting point for which is to model a galaxy's surface brightness by a sum of two-dimensional elliptical Gaussians. \citet{Bendinelli1991} used MGEs for the reconstruction of galaxy images, but only for spherical systems. \citet{Monnet1992} generalized MGEs to triaxial systems and made their application to real galaxies possible. Further development of the formalism and its application was undertaken by \citet{Emsellem1994a,Emsellem1994b}. \citet{Cappellari2002} develop a sectors fitting code to perform MGE fits to galaxy images, and this code made MGEs more readily accessible.

The projected surface brightness can be written as a sum of two-dimensional Gaussians
\begin{equation}
    \Sigma(R',\theta')=\sum_{\rm j=1}^{N} \frac{L_j}{2\pi\sigma_j'^2q_j'}\exp \left[ -\frac{1}{2\sigma_j'^2}(x'^2+\frac{y'^2}{q_j'^2}) \right],
\label{MGE}
\end{equation}
with
\begin{equation}
\left\{
\begin{array}{ll}
    x'=R'\sin(\theta'-\psi),&\\
    y'=R'\cos(\theta'-\psi),
\end{array}
\right.
\end{equation}
where $N$ is the number of Gaussians, the subscript $j$ means the value for $j$-th Gaussian, $L_j$ is the total luminosity, $q_j'=b_j'/a_j'$ is the axis ratio and $\sigma_j'$ is the dispersion along major axis. $(R',\theta')$ are the polar coordinates on the plane of the sky $(x',y')$. $\psi$ is the position angle (PA), which is the angle between the Gaussian major axis and $y'$ axis, measured counterclockwise. Here we note that we do not have subscript $j$ in $x$ and $y$ as Equation (1) in \citet{Cappellari2002} and Equation (4) in \citet{RvdB2008}, because we assume the Gaussians are concentric with their major axes aligned. By using the deprojection method given in \citet{Emsellem1994a}, we can obtain the three-dimensional luminosity distribution (see Equation 8 in \citealp{Emsellem1994a})
\begin{equation}
    \nu(x,y,z)=\sum_{\rm j=1}^{N} \frac{L_j}{(\sqrt{2\pi}\sigma_j)^3p_j q_j}\exp \left[ -\frac{1}{2\sigma_j^2}(x^2+\frac{y^2}{p_j^2}+\frac{z^2}{q_j^2}) \right],
\label{MGE3}
\end{equation}
where $p_j=b_j/a_j$ and $q_j=c_j/a_j$ are the axis ratios for the ellipsoid ($a_j\ge b_j\ge c_j$), while other terms have the same meaning as Equation~(\ref{MGE}). The combination between Equation~(\ref{MGE}) and Equation~(\ref{MGE3}) are described in $\S$~\ref{sec4.4}. We use a constant stellar mass to light ratio $M_*/L$ so that the stellar mass density is given by
\begin{equation}
    \rho_{\rm star}(x,y,z)=(M_*/L)\times\nu(x,y,z).
\end{equation}
The ratio $M_*/L$ is a free parameter in our modelling. The corresponding gravitational potential $\Phi_{\rm star}$ is calculated based on the classical \citet{Chandrasekhar1969} formula (see $\S 3.8$ in \citealp{RvdB2008} for details).

\subsection{NFW dark matter halo}
\label{sec4.2}
We use the spherical Navarro-Frenk-White (NFW) profile described in \citet{Navarro1996} for creating our dark matter haloes. Using the profile, the density of dark matter can be written as
\begin{equation}
    \rho_{\rm DM}(r)=\frac{\rho_0}{\frac{r}{R_{\rm s}} \left( 1+\frac{r}{R_{\rm s}} \right)^2 },
\end{equation}
where $\rho_0$ and the ``scale radius'' $R_{\rm s}$ are two free parameters. By solving Poisson's equation, the potential of NFW profile is

\begin{equation}
    \Phi_{\rm DM}(r)=-\frac{4\pi G\rho_0 R_s^3}{r}\ln(1+\frac{r}{R_s}).
\end{equation}

The virial radius $R_{200}$ is defined as the radius within which the average density is 200 times the critical density $\rho_{\rm crit}$. As the total enclosed mass within $R_{200}$ is $M_{200}$, we have
\begin{equation}
    \frac{M_{200}}{R_{200}^3\times4/3\pi}=200\rho_{\rm crit},
\label{NFW1}
\end{equation}
with
\begin{equation}
    \rho_{\rm crit}=\frac{3H_0^2}{8\pi G},
\end{equation}
where $H_0$ is the Hubble constant and $G$ is the gravitational constant.

From the definition of density profile in Equation~\ref{NFW1}, we obtain
\begin{equation}
    M_{200}=\int_{\rm 0}^{R_{200}}4\pi r^2 \rho(r)dr=4\pi \rho_0 R_{\rm s}^3 \left[ \ln(\frac{R_s+R_{200}}{R_s})-\frac{R_{200}}{R_s+R_{200}} \right].
\end{equation}
Instead of $\rho_0$ and $R_{\rm s}$, we use the ``concentration parameter'' $c$ and ``the fraction of dark matter within $R_{\rm 200}$'' $f$ to represent a NFW halo
\begin{equation}
\left\{
\begin{array}{ll}
    c=R_{200}/R_{\rm s},&\\
    f=M_{200}/M_*,
\end{array}
\right.
\label{NFW2}
\end{equation}
where $M_*$ is the total stellar mass.

Based on equations ~\ref{NFW1} to ~\ref{NFW2}, the parameters in the NFW profile can be rewritten as
\begin{equation}
    \rho_0=\frac{200}{3}\frac{c^3}{\ln(1+c)-c/(1+c)}\times\rho_{\rm crit},
\end{equation}
\begin{equation}
    R_{\rm s}=\left[ \frac{3}{800\pi}\frac{M_*f}{\rho_{\rm crit}c^3} \right]^{1/3}.
\end{equation}
The two parameters $c$ and $f$ of the NFW profile are free parameters in our modelling.
\subsection{Black hole}
\label{sec4.3}
We assume a central black hole can be represented by a Plummer potential \citep{RvdB2008}:
\begin{equation}
    \Phi_{\rm BH}(x,y,z)=-\frac{GM_{\rm BH}}{\sqrt{r_{\rm soft}^2+x^2+y^2+z^2}},
\end{equation}
where $r_{\rm soft}$ is the softening length, which is introduced to prevent the central potential to be infinite, and equals to 0.001 arcsec in our modelling runs. All our galaxies have central black holes but, given the nature of our research, they will not strongly influence our results. We choose therefore to only use a token black hole which also helps in
reducing our computing costs. We therefore fix the black hole mass $M_{\rm BH}=10^6M_{\odot}$ for all mock data sets.
\subsection{Intrinsic shapes and viewing angles}
\label{sec4.4}
A two-dimensional Gaussian has two further parameters in addition to luminosity: the deviation along major axis $\sigma_j'$ and the axis ratio $q_j'$. For a three-dimensional Gaussian, it becomes $\sigma_j$ and axis ratios $(p_j,q_j)$. Here we use the compression factor $u_j=\sigma_j'/\sigma_j$ to present the difference between the length of the intrinsic major axis and the major axis on the projected plane. For a triaxial galaxy, we set the major axis to be the $x-$axis in Cartesian coordinates. There are three viewing angles to be considered, angles in spherical coordinates $(\theta,\varphi)$ and the rotation $\psi$ of the projected major axis on the sky-plane, measured counterclockwise from the $y'-$axis. The original coordinates $(x,y,z)$, the line of sight coordinates $(x',y',z')$ and the projected coordinates after rotation $(x'',y'',z'')$ are related by
\begin{equation}
\begin{bmatrix}
    x''\\
    y''\\
    z''
\end{bmatrix}
=
\begin{bmatrix}
    \sin\psi & -\cos\psi & 0\\
    \cos\psi & \sin\psi  & 0\\
    0        & 0         & 1
\end{bmatrix}
\begin{bmatrix}
    x'\\
    y'\\
    z'
\end{bmatrix}
,
\end{equation}

\begin{equation}
\begin{bmatrix}
    x'\\
    y'\\
    z'
\end{bmatrix}
=
\begin{bmatrix}
    -\sin\varphi           & \cos\varphi            & 0\\
    -\cos\theta\cos\varphi & -\cos\theta\sin\varphi & \sin\theta\\
    \sin\theta\cos\varphi  & \sin\theta\sin\varphi  & \cos\theta
\end{bmatrix}
\begin{bmatrix}
    x\\
    y\\
    z
\end{bmatrix}
.
\end{equation}

After assuming the viewing angles $(\theta,\varphi,\psi)$ in models, we can covert observed quantities $(\sigma_j',q_j')$ into intrinsic parameters $(\sigma_j,p_j,q_j)$, based on formulas below (\citealp{RvdB2008}, \citealp{Cappellari2002})
\begin{equation}
    1-q_j^2=\frac{\delta_j'[2\cos2\psi+\sin2\psi(\sec\theta\cot\varphi-\cos\theta\tan\varphi)]}{2\sin^2\theta[\delta_j'\cos\psi(\cos\psi+\cot\varphi\sec\theta\sin\psi)-1]},
\end{equation}

\begin{equation}
    p_j^2-q_j^2=\frac{\delta_j'[2\cos2\psi+\sin2\psi(\cos\theta\cot\varphi-\sec\theta\tan\varphi)]}{2\sin^2\theta[\delta_j'\cos\psi(\cos\psi+\cot\varphi\sec\theta\sin\psi)-1]},
\end{equation}

\begin{equation}
    u_j^2=\frac{1}{q_j'^2}\sqrt{p_j^2\cos^2\theta+q_j^2\sin^2\theta(p_j^2\cos^2\varphi+\sin^2\varphi)},
\end{equation}
where $\delta_j'=1-q_j'^2$.

In reverse, if we know the intrinsic shapes $(p,q)$ combined with the compression factor $u$ of any deprojected Gaussian along with the projected quantity $q'$, we can obtain the space orientation parameters $(\theta,\varphi,\psi)$
\begin{equation}
    \cos^2\theta=\frac{(u^2-q^2)(q'^2u^2-q^2)}{(1-q^2)(p^2-q^2)},
\end{equation}
\begin{equation}
    \tan^2\varphi=\frac{(u^2-p^2)(p^2-q'^2u^2)(1-q^2)}{(1-u^2)(1-q'^2u^2)(p^2-q^2)},
\end{equation}
\begin{equation}
    \tan^2\psi=\frac{(1-q'^2u^2)(p^2-q'^2u^2)(u^2-q^2)}{(1-u^2)(u^2-p^2)(q'^2u^2-q^2)},
    \label{transformation}
\end{equation}
where the parameters should satisfy $q \le p \le 1$, $q \le q'$ and $\max(q/q',p) \le u \le \min(p/q',1)$ for validation of the equation.

In the modelling, we take $(p,q,u)$ as free parameters instead of $(\theta,\varphi,\psi)$ as they directly represent the intrinsic morphologies of galaxies. We use $p$ and $q$ in our definition of triaxiality \citep{Binney2008} and take
\begin{equation}
    T=\frac{1-p^2}{1-q^2}.
\end{equation}

\subsection{Orbit distribution parameters}
\label{sec4.5}
We introduce three parameters to characterize orbits, time-averaged radius $r$, circularity $\lambda_z$ and $\lambda_x$. We define orbit circularity $\lambda_z$ as in \citet{Zhu2018a,Zhu2018b} as a ratio of time-averaged quantities
\begin{equation}
    \lambda_z=\overline{L_z}/(\overline{r}\times\overline{V_{\rm rms}}),
\end{equation}
where $\overline{L_z}=\overline{xv_y-yv_x}$, $\overline{r}=\overline{\sqrt{x^2+y^2+z^2}}$ and $\overline{V_{\rm rms}}=\sqrt{\overline{v_x^2+v_y^2+v_z^2+2v_xv_y+2v_xv_z+2v_yv_z}}$. The numerator is the the time averaged z-component of the the orbit's angular momentum, and the denominator is the product of two time-averaged root mean square (rms) quantities calculated from the orbit, the average radial position and velocity. If we treat the rms velocity as a circular velocity in the equatorial plane, then the product represents the angular momentum of a typical circular orbit associated with the original orbit. The ratio of the two angular momentum terms gives us our measure of orbit circularity. Note that it is the circularity distribution as a function of radius which is important for comparison purposes. Whilst there is a physical explanation for our orbit circularity definition, it can not be justified precisely mathematically, and should be treated as empirical. Thus $\lambda_z\sim 1$ represents highly rotating short-axis tube orbits, while $\lambda_z\sim 0$ is mostly long-axis tube and box orbits. Since there are prolate-like galaxies in our sample and $\lambda_z$ can not distinguish long-axis tube orbits from box orbits, we also define $\lambda_x$, which represents the normalized angular momentum around the major axis ($x$-axis)
\begin{equation}
    \lambda_x=\overline{L_x}/(\overline{r}\times\overline{V_{\rm rms}}),
\end{equation}
where $\overline{L_x}=\overline{yv_z-zv_y}$. The orbit circularity distribution can be thought of the probability densities of orbits on the coordinate space $\lambda_z$ versus $r$ or $\lambda_x$ versus $r$.

Velocity dispersion anisotropy parameters are widely used as indicators of the underlying orbit distribution of a galaxy. The velocity anisotropy $\beta_r$ \citep{Binney2008} is defined as
\begin{equation}
    \beta_r=1-\frac{\overline{\sigma_{\rm \theta}^2}+\overline{\sigma_{\rm \varphi}^2}}{2\overline{\sigma_r^2}}\equiv1-\frac{\overline{\sigma_t^2}}{2\overline{\sigma_r^2}},
\end{equation}
where $\sigma_r$, $\sigma_{\theta}$ and $\sigma_{\varphi}$ are the three components of the velocity dispersion in spherical coordinates, and $\sigma_t^2=\sigma_{\theta}^2+\sigma_{\varphi}^2$. $\beta_r>0$ indicates radial anisotropy and $\beta_r<0$ indicates tangential anisotropy. However, $\beta_r$ ranges from negative infinity to 1, and is not uniformly distributed in the meaning. In order to overcome this shortcoming of $\beta_r$ for quantitative analysis, we define a new anisotropy parameter the ``tangential fraction'' ($0\le f_t\le1$), as
\begin{equation}
    f_t= \frac{(\overline{\sigma_{\rm \theta}^2}+\overline{\sigma_{\rm \varphi}^2})/2}{\overline{\sigma_r^2}+(\overline{\sigma_{\rm \theta}^2}+\overline{\sigma_{\rm \varphi}^2})/2}\equiv\frac{\overline{\sigma_t^2}}{2\overline{\sigma_r^2}+\overline{\sigma_t^2}}.
    \label{tangential}
\end{equation}
$f_t>0.5$ represents tangential anisotropy while $f_t<0.5$ represents radial anisotropy, and $f_t=0.5$ is for isotropy.


\subsection{$\chi^2$ and the best-fitting model}
\label{sec4.6}
We have a total of six free parameters when we run models. They are intrinsic shape parameters $p$ , $q$ and $u$ ($\S$~\ref{sec4.4}), NFW dark matter halo parameters $c$ and $f$ ($\S$~\ref{sec4.2}) and the stellar mass to light ratio $M_*/L$ ($\S$~\ref{sec4.1}). We use both kinematic data and luminosity distributions as constraints. The $\chi_{\rm NNLS}^2$ function that needs to be minimized by NNLS contains two components:
\begin{equation}
    \chi_{\rm NNLS}^2=\chi_{\rm lum}^2+\chi_{\rm kin}^2.
\label{chi2NNLS}
\end{equation}

For the luminous component, we use both the two-dimensional surface brightness $S_n$ ($n$-th Voronoi bin in the observing plane) and the three-dimensional luminosity distribution $\rho_m$ ($m$-th Schwarzschild cell in 3D space) as constraints. Following \citep{Zhu2018a}, we set relative errors of $S_n$ to be $1\%$ and $\rho_n$ to be $2\%$, thus
\begin{equation}
    \chi_{\rm lum}^2=\sum_{\rm n=1}^{N}\left(\frac{S_n^*-S_n}{0.01S_n}\right)^2+\sum_{\rm m=1}^{M}\left(\frac{\rho_n^*-\rho_n}{0.02\rho_n}\right)^2,
\end{equation}
where the variables marked with a ``$*$'' indicates model predictions and those without a ``$*$'' are observational data which are constructed from the MGE formulae.

For kinematic components, we use Gauss-Hermite series \citep{vdM1993,Gerhard1993,Rix1997} to fit the velocity distribution profile in each bin. Considering Gauss-Hermite coefficients $h_1$ and $h_2$, we have
\begin{equation}
    \chi_{\rm kin}^2=\sum_{\rm n=1}^{N}\left[ \left(\frac{h^*_{1n}-h_{1n}}{\Delta h_{1n}}\right)^2+\left(\frac{h^*_{2n}-h_{2n}}{\Delta h_{2n}} \right)^2\right],
\end{equation}
where the markers ``$\Delta$'' represent observational errors. The kinematics from the model are luminosity-weighted in the same way as the observations. After modelling, $h_1$, $h_2$, $\Delta h_1$ and $\Delta h_2$ will be converted to $V$, $\sigma$, $\Delta V$ and $\Delta \sigma$ for comparison purposes \citep{Zhu2018a}.

We explore the parameter space by changing the values of our free parameters and reconstructing our gravitational potential accordingly. The modelling begins from assumed start parameters and iterates towards models with smaller chi-square values in fixed steps. The differences between model values and observed values are calculated as
\begin{equation}
    \chi^2=\sum_{\rm n=1}^{N}\left[ \left(\frac{V^*_n-V_n}{\Delta V_n}\right)^2+\left(\frac{\sigma^*_n-\sigma_n}{\Delta \sigma_n} \right)^2\right],
\label{chi2kin}
\end{equation}
where $V^*_n$, $\sigma^*_n$ are model values for $n$-th bin and $V_n$, $\sigma_n$ mean observed values, while $\Delta V_n$, $\Delta \sigma_n$ represent observational errors. We use $\chi^2$ (Equation~\ref{chi2kin}) to search for the best-fitting model. Considering Equation~\ref{chi2NNLS}, $\chi^2_{\rm lum}$ is usually small, $\chi^2$ is in fact strongly correlated with $\chi^2_{\rm NNLS}$. Typically, a completed model with six free parameters require $1000\sim2000$ Schwarzschild modelling runs depending on how close the start parameter values are to the best-fitting model. We normalize $\chi^2$ by forcing $\chi^2_{\rm min}=N_{\rm obs}$, thus the normalized $\chi'^2=\chi^2/\chi^2_{\rm min}\times N_{\rm obs}$.

In \citet{RvdB2008}, the $1\sigma$ confidence level is defined as
\begin{equation}
    \Delta\chi'^2 \equiv \chi'^2-\chi'^2_{\rm min}<\sqrt{2(N_{\rm obs}-N_{\rm par})},
\end{equation}
where $N_{\rm obs}=2\times N_{\rm kin}$ when we only use $V$ and $\sigma$ as model constraints. $N_{\rm par}$ means the number of free parameters and equals to 6 here, while $N_{\rm kin}$ is usually around several hundreds or one thousand in our mock data sets.

We can see with the increase of $N_{\rm obs}$, $\Delta\chi'^2/\chi'^2$ goes down, which means $\sqrt{2(N_{\rm obs}-N_{\rm par})}$ could be smaller than the model $\chi'^2$ fluctuation thus underestimate the $1\sigma$ region when $N_{\rm obs}$ is large. In order to solve this problem, we try to re-scale $1\sigma$ confidence level by
\begin{equation}
    \Delta\chi'^2 \equiv \chi'^2-\chi'^2_{\rm min}<A\times\sqrt{2(N_{\rm obs}-N_{\rm par})},
\label{chi2}
\end{equation}
where $A$ is a constant parameter to be determined. For each mock galaxy, we perturb the kinematic maps $(V_i,\sigma_i)$ with their error maps $(\Delta V_i,\Delta \sigma_i)$: $V'_i=V_i+\Delta V_i\times a$, $\sigma'_i=\sigma_i+\Delta \sigma_i\times b$, where $a$, $b$ are random values with standard deviation equals unity. By fixing the gravitational potential, we re-solve the orbits weight with NNLS, and its result $\chi'^2_{\rm perturb}$ is slightly different with $\chi'^2$. We do the perturbations and find orbit weight solutions for 100 times. The standard deviation of these 100 $\chi'^2_{\rm perturb}$ is taken as the model $\chi'^2$ fluctuation. We find the parameter $A$ ranges from $1.03$ to $3.19$ for different mock galaxies, with an average of $2.01$ (this is consistent with the CALIFA model fluctuations in \citealp{Zhu2018a}). Thus we roughly take $A=2$ and Equation~\ref{chi2} become
\begin{equation}
    \Delta\chi'^2 \equiv \chi'^2-\chi'^2_{\rm min}<2\times\sqrt{2(N_{\rm obs}-N_{\rm par})},
\label{1sigma}
\end{equation}
which will be used as the $1\sigma$ confidence level in our models constrained by MaNGA/CALIFA quality data.

The $1\sigma$ confidence level we define here represents the model fluctuations, and we take it as the modelling error in our results analysis. A standard $\chi^2$ analysis is not valid here for two reasons. First, for our models, the number of active orbits could be larger than the number of data points used as model constraints, so that the degrees of freedom $N_{\rm kin}-N_{\rm orb}$ in a $\chi^2$ distribution will be negative \citep{Press1992}. Second, the typical $\Delta \chi^2$ of the standard $\chi^2$ analysis is equal to 7.04 (joint constraints on 6 free parameters simultaneously) for 1 $\sigma$ confidence level, while the model fluctuation is relatively large ($\Delta \chi^2 \sim 130$ for $N_{\rm obs}\sim 2000$). That means if we use the standard $\chi^2$ analysis, $\Delta \chi^2$ (errors of the model) will be significantly under-estimated. Using model fluctuations as confidence levels is an empirical way of calculating the error bars. We refer the readers to \citet{Morganti2013}, who applied the M2M method to NGC 4494, for detailed discussions on the approach and justifications.

\section{Simulated galaxies and Mock data}
\label{sec5}
\subsection{Simulated galaxies}
\label{sec5.1}
We use the highest resolution simulation, Illustris-1, in the Illustris project \citep{Vogelsberger2014a,Vogelsberger2014b,Genel2014,Nelson2015}, which contains $2\times1820^3$ resolution elements in a volume of $(106.5\ \rm Mpc)^3$, with $6.26\times10^6 M_{\odot}$ dark matter mass resolution and an initial baryonic mass resolution of $1.26\times10^6 M_{\odot}$.

We select nine early-type Illustris galaxies with stellar mass ranging from $10^{10.74}$ to $10^{12.05}M_{\odot}$. Since effective radius $R_{\rm e}$ is usually calculated from a galaxy's surface brightness, its value may change with viewing angles. We use the average effective radius of five different projections $\overline{R_{\rm e}}$ as the typical size of a galaxy. We calculate the intrinsic shapes of these galaxies by using the reduced inertia tensor method \citep{Allgood2006}. For the stellar components, these nine galaxies have different morphologies and comprise three oblate galaxies subhalo73666, subhalo347122, subhalo336920 (galaxy O1, O2, O3 for short) with their triaxial parameter values at $\overline{R_{\rm e}}$, $T_{\rm e}$, typically $\le0.3$; three triaxial galaxies subhalo312924, subhalo175435, subhalo16942 (galaxy T1, T2, T3 for short) with $0.3<T_{\rm e}<0.7$; and three prolate galaxies subhalo73664, subhalo191220, subhalo222715 (galaxy P1, P2, P3 for short) with $T_{\rm e}\ge0.7$. We refer to them as oblates, triaxials and prolates. The dark matter haloes in these galaxies are triaxial with short to long axis ratios between 0.63 and 0.92, and medium to long axis ratios between 0.70 and 0.95 at 30 kpc. Detailed information on the sample galaxies is shown in Table~\ref{galaxy catalog}.
\subsection{Mock data}
\label{sec5.2}

\begin{figure*}
\begin{centering}
	\includegraphics[width=10cm]{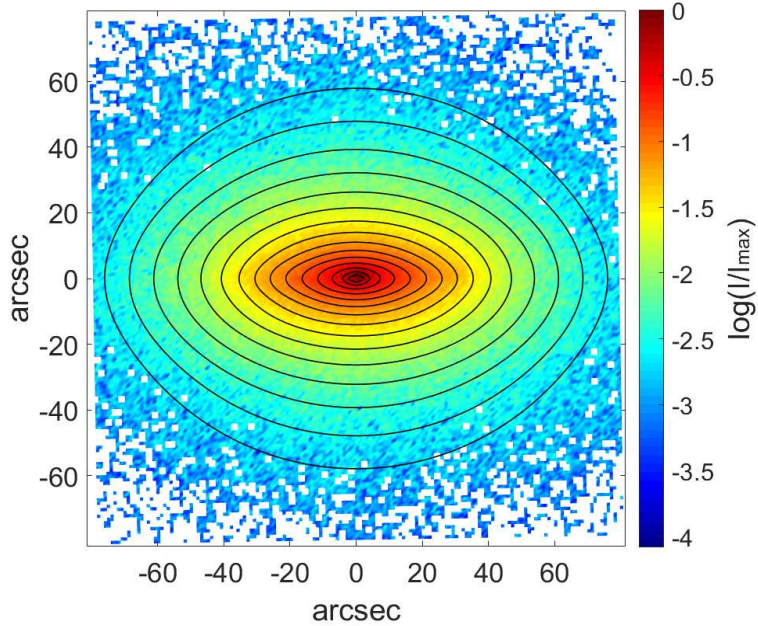}
    \caption{The distribution of surface brightness (projected stellar mass) and MGE fitting contours of galaxy O1 with viewing angles $(\theta,\varphi)=(83,131)^{\circ}$. The color map shows the distribution of surface brightness, with the relative intensity $\log(I/I_{\rm max})$ indicated by the color bar. The black contours are the luminous contours obtained from MGE fitting method, with an interval of 0.5 magnitude.}
    \label{img}
\end{centering}
\end{figure*}
\begin{figure*}
\begin{centering}
	\includegraphics[width=16cm]{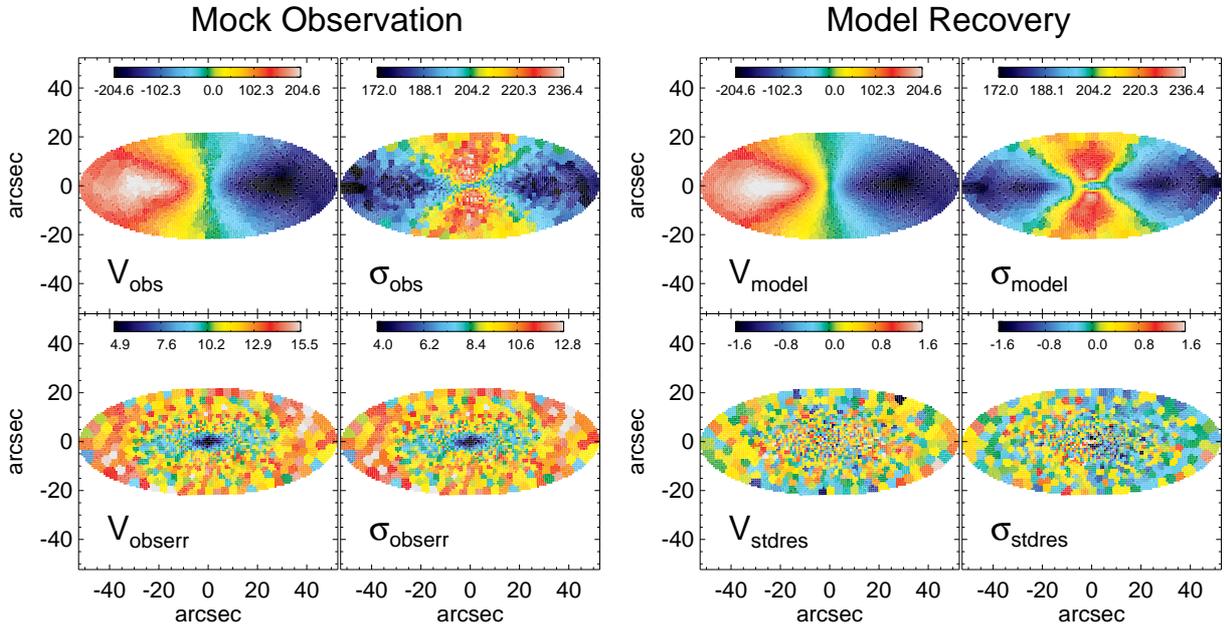}
    \caption{The kinematic mock data set and model recovery of galaxy O1 with viewing angles $(\theta,\varphi)=(83,131)^{\circ}$. Left: the Voronoi-binned kinematic map of mock observation, including line of sight mean velocity $V_{\rm obs}$ (top left), velocity dispersion $\sigma_{\rm obs}$ (top right) and their observational error $V_{\rm obserr}$ (bottom left), $\sigma_{\rm obserr}$ (bottom right). Each panel has an independent color bar to show the data values. Right: the model recovery of mean velocity $V_{\rm model}$ (top left), velocity dispersion $\sigma_{\rm model}$ (top right) together with the standardized residuals between the inputs and outputs $V_{\rm stdres}=(V_{\rm model}-V_{\rm obs})/V_{\rm obserr}$, $\sigma_{\rm stdres}=(\sigma_{\rm model}-\sigma_{\rm obs})/\sigma_{\rm obserr}$.}
    \label{mock-data}
\end{centering}
\end{figure*}

The Illustris galaxies contain both stellar particles and dark matter particles with a resolution of $\sim 10^6 M_{\odot}$, as mentioned in $\S$~\ref{sec5.1}. We project the stellar particles in each galaxy onto five different observational planes by setting specific viewing angles. The angles $(\theta,\varphi)$ are randomly chosen in spherical coordinates, with probability $dP=d\Omega/4\pi=\sin\theta d\theta d\varphi/4\pi$, where $\Omega$ represents the solid angle. Due to symmetry, we restrict $0\le\theta\le\pi/2$ and $0\le\varphi<\pi$. The angles $(\theta,\varphi)$ we have chosen for the mock data are shown in Table~\ref{galaxy catalog}. We create surface brightness and kinematic maps, which are used to generate model constraints, for each simulated galaxy with the chosen viewing angles $(\theta,\varphi)$. Making allowances for spatial differences, the way we generate luminosity input and kinematic input follows the method in \citet{Li2016}. We create $5\times9=45$ mock data sets in total, and treat each as an independent observed galaxy in the modelling.

The surface mass density $\Sigma^*(n)$ is calculated with a pixel size of 0.5kpc$\times$0.5kpc. Instead of using the luminosity distribution given in the simulation directly, we use the stellar mass distribution to construct mock galaxies with a constant stellar mass-to-light ratio $M_*/L=5$. We can obtain the surface brightness $\Sigma(n)=\Sigma^*(n)/5$ and fit it with the MGE method to generate the luminosity input (see $\S$~\ref{sec4.1}). Finally, we rotate the major axis of surface brightness to be horizontal in the observational plane, ensuring the other viewing angle $\psi\approx90^{\circ}$ (see $\S$~\ref{sec4.4}).

Kinematic data can be obtained from the velocity distribution of stellar particles, also in a 0.5 kpc$\times$0.5 kpc pixels. In this paper, we use the luminosity-weighted mean velocity $V$ and velocity dispersion $\sigma$ (standard deviation of line-of-sight velocity) together with their observational errors, converted to Gauss-Hermite coefficients $h_1$ and $h_2$, as constraints. In order to achieve consistency in approach between the simulation tests and real observations, we do not model with skewness $h_3$ and kurtosis $h_4$ as they are not always able to be determined well in real IFU surveys like MaNGA. We bin the pixels to nearly constant signal-to-noise ratio $S/N=20$ per bin using a Voronoi 2D-binning method (\citealp{Cappellari2003}). By projecting the stellar particles onto the observing plane and obtaining the line of sight velocities of stellar particles located in each bin, we calculate the luminosity-weighted mean velocity $V$ and velocity dispersion $\sigma$ by fitting the binned data with a Gaussian. Observational errors are estimated by using the bootstrap method (random sampling with replacement). We randomly select half the number of the stellar particles in each bin with replacement 500 times, and calculate $V_1,V_2,...V_{500},\sigma_1,\sigma_2,...,\sigma_{500}$ for our selections. The standard deviation of these 500 mean velocities and velocity dispersions are taken as observational errors $\Delta V$ and $\Delta \sigma$. As triaxial models are point-symmetric, we lastly point-symmetrize the kinematic data in the Voronoi bins.

We take galaxy O1 with viewing angles $(\theta,\varphi)=(83,131)^{\circ}$ as an example in Figure~\ref{img} and Figure~\ref{mock-data}. Figure~\ref{img} shows the surface brightness distribution and the contours obtained from MGE fitting. Figure~\ref{mock-data} presents the kinematic inputs and show the observational mean velocity $V_{\rm obs}$, and velocity dispersion $\sigma_{\rm obs}$ together with their errors $V_{\rm obserr}$, $\sigma_{\rm obserr}$. For ease of comparsion we also show the model estimated $V_{\rm model}$, $\sigma_{\rm model}$ together with the standardized residuals between the inputs and outputs $V_{\rm stdres}=(V_{\rm model}-V_{\rm obs})/V_{\rm obserr}$, $\sigma_{\rm stdres}=(\sigma_{\rm model}-\sigma_{\rm obs})/\sigma_{\rm obserr}$. Note that we use units of arcsec in Figure~\ref{img} and Figure~\ref{mock-data} by assuming the galaxy is located at the redshift of $z=0.03$.

\begin{table*}
\caption{The catalog of nine simulated galaxies. From top to bottom are: (1) the galaxy morphology; (2) galaxy name (Illustris object number); (3) galaxy short name used in our paper; (4) total stellar mass $\log(M_*/M_{\odot})$; (5) central black hole mass $\log(M_{\rm BH}/M_{\odot})$; (6) number of stellar particles $N_{\rm star}$; (7) number of dark matter particles $N_{\rm dark}$; (8) total gas fraction $f_{\rm gas}$; (9) average effective radius $\overline{R_{\rm e}}$; (10) maximum data coverage ${R_{\rm m}}$; (11) axis ratios $p_{\rm e}$, $q_{\rm e}$ and triaxial parameter $T_{\rm e}$ at $\overline{R_{\rm e}}$; (12) viewing angles $\theta$ and $\varphi$ of five random projections. There are totally $5\times9=45$ mock data sets created.}
    \centering
	\begin{tabular}{|c|c|c|c|c|c|c|c|c|c|c|}
    \hline
    \multicolumn{2}{|c|}{Morphology} & \multicolumn{3}{|c|}{Oblates} & \multicolumn{3}{|c|}{Triaxials} & \multicolumn{3}{|c|}{Prolates}\\
	\hline
	\multicolumn{2}{|c|}{Galaxy name (subhalo)} & 73666 & 347122 & 336920 & 312924 & 175435 & 16942 & 73664 & 191220 & 222715\\
	\multicolumn{2}{|c|}{Short name} & O1 & O2 & O3 & T1 & T2 & T3 & P1 & P2 & P3\\
	\hline
	\multicolumn{2}{|c|}{$\log(M_*/M_{\odot})$} & 11.76 & 11.21 & 10.74 & 11.56 & 11.61 & 11.54 & 12.05 & 11.99 & 11.72\\
	\multicolumn{2}{|c|}{$\log(M_{\rm BH}/M_{\odot})$} & 9.02 & 8.58 & 7.71 & 8.73 & 9.34 & 8.91 & 9.84 & 9.65 & 9.30\\
	\hline
	\multicolumn{2}{|c|}{$N_{\rm star}$ ($10^5$)} & 7.01 & 2.04 & 0.65 & 4.52 & 5.01 & 4.35 & 13.98 & 12.14 & 6.55\\
	\multicolumn{2}{|c|}{$N_{\rm dark}$ ($10^5$)} & 13.58 & 6.18 & 4.82 & 8.04 & 35.39 & 3.99 & 48.76 & 56.26 & 23.12\\	
	\multicolumn{2}{|c|}{$f_{\rm gas}$ ($\%$)} & 0.19 & 0.17 & 1.36 & 0.40 & 0.31 & 0.00 & 0.27 & 0.14 & 0.13\\
	\hline
	\multicolumn{2}{|c|}{$\overline{R_{\rm e}}$ (kpc)} & 10.46 & 7.06 & 5.12 & 7.42 & 7.06 & 9.18 & 8.96 & 16.09 & 9.30\\
	\multicolumn{2}{|c|}{$R_{\rm m}$ (kpc)} & 32.30 & 22.36 & 15.10 & 22.82 & 18.46 & 28.83 & 24.56 & 50.56 & 31.23\\
    \hline
    \multirow{3}*{Intrinsic shapes} & $p_{\rm e}$ & 0.98 & 0.89 & 0.92 & 0.80 & 0.76 & 0.68 & 0.71 & 0.69 & 0.52\\
	~                               & $q_{\rm e}$ & 0.44 & 0.34 & 0.59 & 0.42 & 0.55 & 0.45 & 0.63 & 0.61 & 0.50\\
	~                               & $T_{\rm e}$ & 0.04 & 0.23 & 0.24 & 0.43 & 0.59 & 0.67 & 0.81 & 0.83 & 0.97\\
	\hline
	\multirow{5}*{Viewing angles} & \multirow{5}*{$(\theta,\varphi)$} & $(17,87)^{\circ}$ & $(27,78)^{\circ}$ & $(10,58)^{\circ}$ & $(40,119)^{\circ}$ & $(28,107)^{\circ}$ & $(3,77)^{\circ}$ & $(36,71)^{\circ}$ & $(18,19)^{\circ}$ & $(18,103)^{\circ}$\\
	~ & ~ & $(42,60)^{\circ}$ & $(53,79)^{\circ}$ & $(38,131)^{\circ}$ & $(63,123)^{\circ}$ & $(43,134)^{\circ}$ & $(45,114)^{\circ}$ & $(41,58)^{\circ}$ & $(45,116)^{\circ}$ & $(37,60)^{\circ}$\\
	~ & ~ & $(55,136)^{\circ}$ & $(61,24)^{\circ}$ & $(55,52)^{\circ}$ & $(70,47)^{\circ}$ & $(57,88)^{\circ}$ & $(49,65)^{\circ}$ & $(52,164)^{\circ}$ & $(52,54)^{\circ}$ & $(66,140)^{\circ}$\\
	~ & ~ & $(74,54)^{\circ}$ & $(74,151)^{\circ}$ & $(73,29)^{\circ}$ & $(76,3)^{\circ}$ & $(67,110)^{\circ}$ & $(62,165)^{\circ}$ & $(63,42)^{\circ}$ & $(67,107)^{\circ}$ & $(75,146)^{\circ}$\\
	~ & ~ & $(83,131)^{\circ}$ & $(85,92)^{\circ}$ & $(81,99)^{\circ}$ & $(86,38)^{\circ}$ & $(69,6)^{\circ}$ & $(86,172)^{\circ}$ & $(84,28)^{\circ}$ & $(87,174)^{\circ}$ & $(88,77)^{\circ}$\\
    \hline
    \end{tabular}
    \label{galaxy catalog}
\end{table*}
\section{Results}
\label{sec6}

In this section, we show model estimates of mass distributions and orbit distributions, including enclosed mass profiles in $\S$~\ref{sec6.1}, galaxy morphologies in $\S$~\ref{sec6.2}, circularity distributions $f(r,\lambda_z)$, $f(r,\lambda_x)$ in $\S$~\ref{sec6.3} and anisotropy parameters $(\beta_r,f_t)$ in $\S$~\ref{sec6.4}. In addition, we report on a simple test concerning the influence of the initial conditions $(E,I_2,I_3)$ in $\S$~\ref{sec6.5}.
\subsection{Mass profile}
\label{sec6.1}
\begin{figure*}
\begin{center}
	\includegraphics[width=16cm]{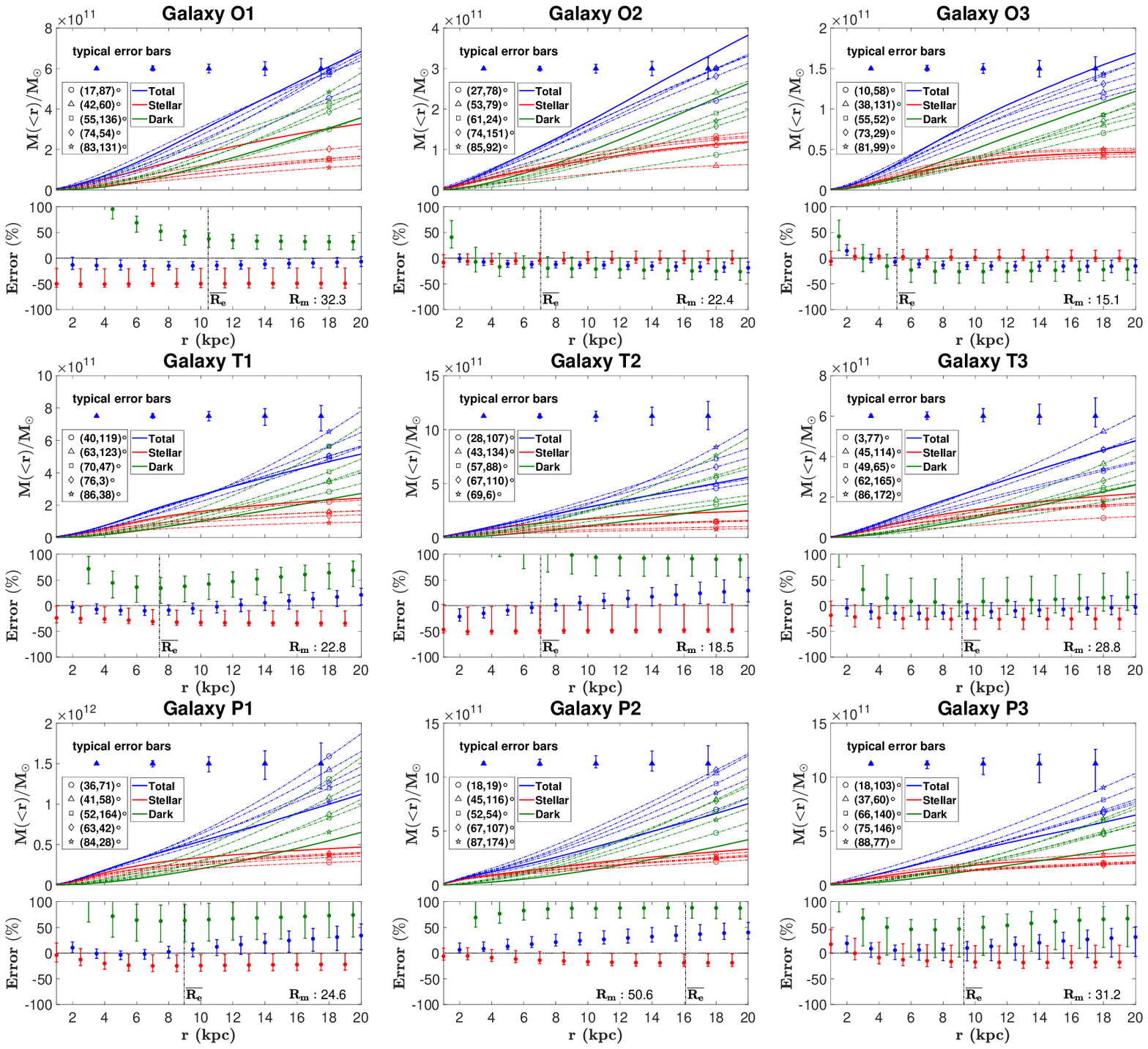}
    \caption{The comparison of the mass profiles $M(\textless r)$ in simulated galaxies and those recovered by our models. Top panels: oblate galaxies O1, O2 and O3. Middle panels: triaxial galaxies T1, T2 and T3. Bottom panels: prolate galaxies P1, P2 and P3. The blue lines and symbols represent the total mass, the red lines and symbols represent the stellar mass, while the green ones are for dark matter. The true profiles are plotted with solid lines, and dashed lines with different markers indicate the model results of different viewing angles $(\theta,\varphi)$. The blue triangles with error bars in the top of each panel represent the typical error bars of model estimated total mass. The dots in the bottom of each panel show the relative average difference $1-\langle M_{\rm true}(\textless r)/M_{\rm model}(\textless r)\rangle$ between true values and model estimated values from mock data with five different projections, and the corresponding error bars represent average confidence levels within $1\sigma$. The vertical dashed line represents $\overline{R_{\rm e}}$ of each galaxy and the value of maximum data coverage $R_{\rm m}$, which means the points outside this value are not reliable, is shown as text.}
    \label{mass}
\end{center}
\end{figure*}
\begin{figure*}
\begin{center}
	\includegraphics[width=16cm]{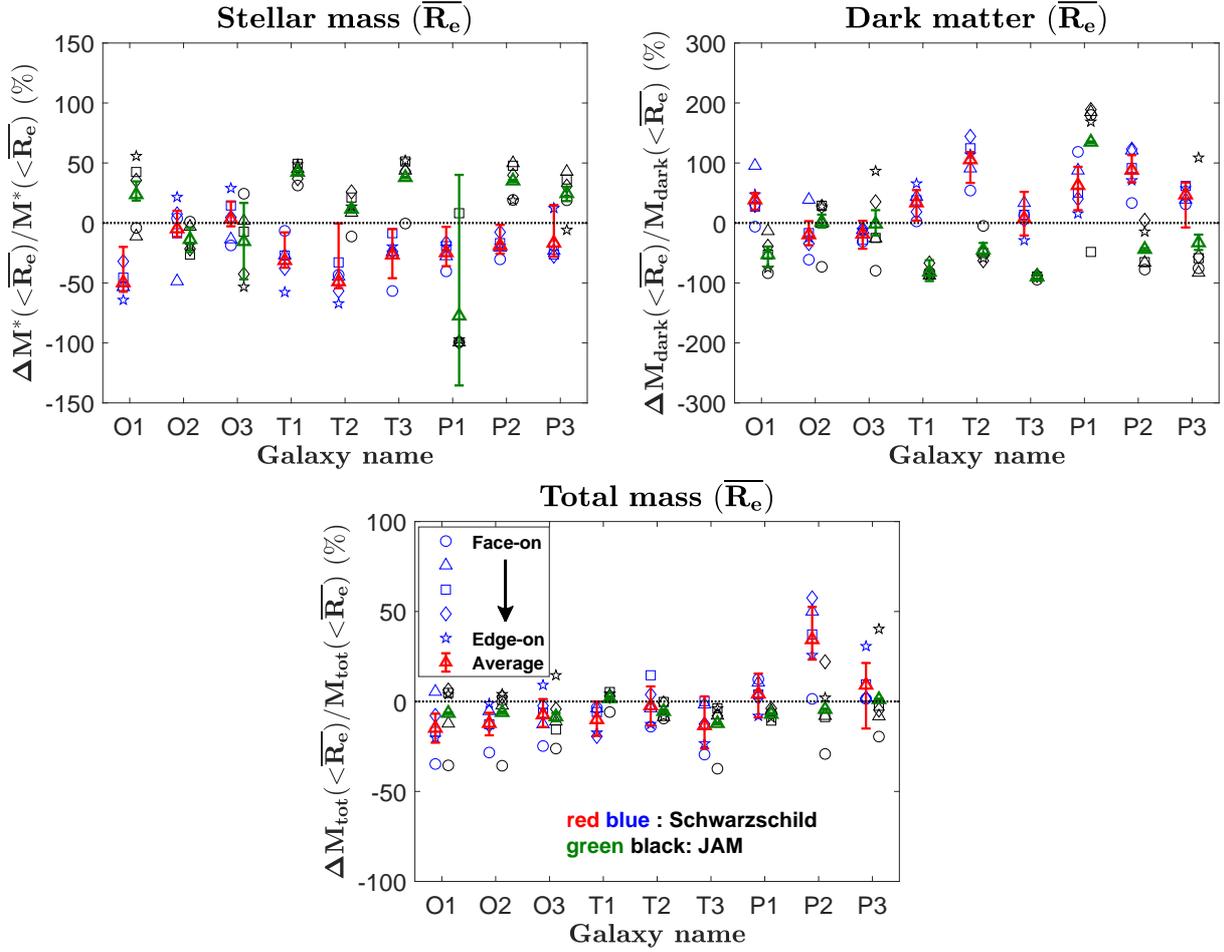}
    \caption{The relative deviation of the enclosed mass between model results and true values. Top panels: stellar mass $\Delta M_*/M_*$ and dark matter mass $\Delta M_{\rm dark}/M_{\rm dark}$ within $\overline{R_{\rm e}}$. Bottom panel: total mass $\Delta M_{\rm tot}/M_{\rm tot}$ within $\overline{R_{\rm e}}$. The horizontal axes indicate galaxy names, from left to right are: oblate galaxies O1, O2 and O3; triaxial galaxies T1, T2 and T3; prolate galaxies P1, P2 and P3. The blue markers represent mock data sets with different viewing angles, from face-on to edge-on are circles, triangles, squares, diamonds, pentagrams. The red triangles with error bars represent the mean values of those obtained from five mock data sets of the same galaxy. The black and green markers show the corresponding results from an oblate JAM technique \citep{Li2016}, using the same mock data sets, but with a generalized NFW halo and more restrictive dynamical assumptions.}
    \label{mass-all}
\end{center}
\end{figure*}
\begin{figure*}
\begin{center}
	\includegraphics[width=16cm]{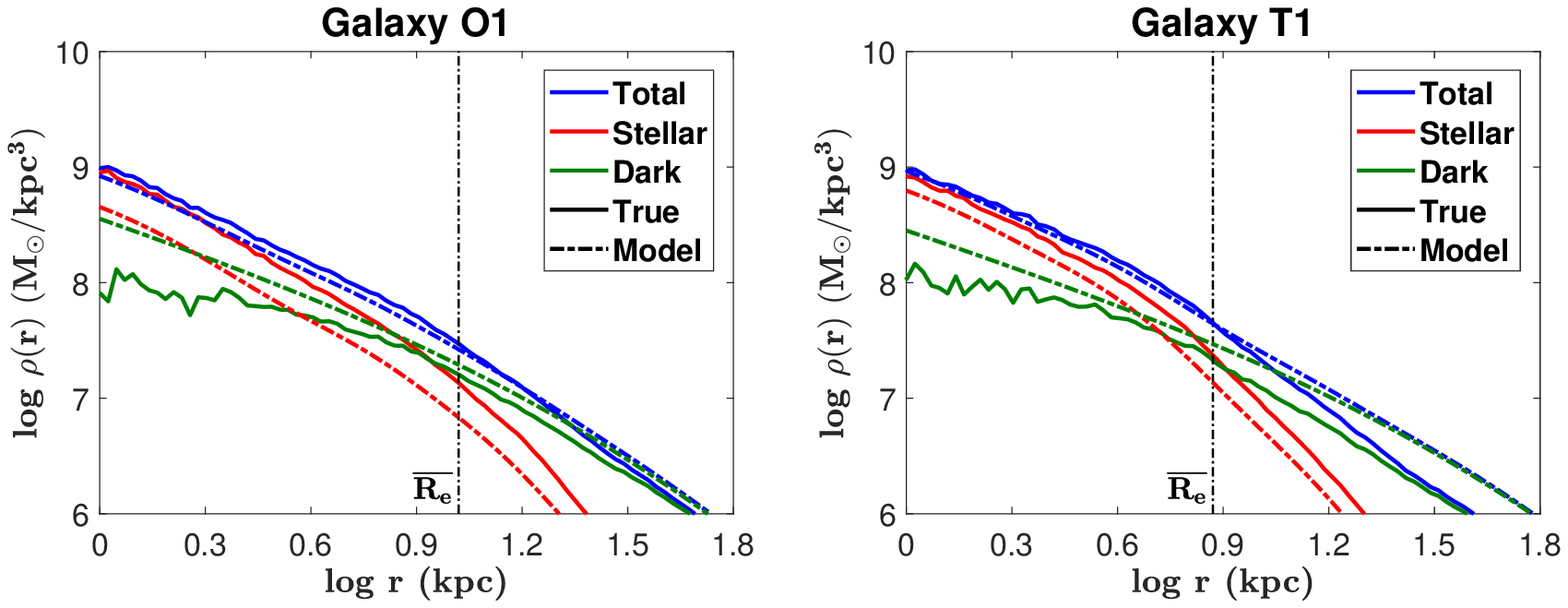}
    \caption{The comparison of the mass density profiles $\rho(r)=\Delta M(r)/(4\pi r^2\Delta r)$ for two example galaxies O1 and T1. The blue lines represent the total mass, the red lines represent the stellar mass, while the green ones are for dark matter (with NFW halo assumed in our model). True profiles are indicated by solid lines while the dashed lines show the mean values of results from mock data with five different projections.}
    \label{mass-density-slope}
\end{center}
\end{figure*}

Mass estimation is the first capability, in line with our objectives, that we consider. We estimate the mass variation with spherical radius $r$ as it is independent of the morphology assumption. Here we compare the total mass $M_{\rm tot}(\textless r)$, the stellar mass $M_*(\textless r)$ and the dark matter mass $M_{\rm dark}(\textless r)$.

Figure~\ref{mass} shows the true mass profile $M(\textless r)$ calculated from the simulation and the profile obtained from modelling. The blue lines and symbols represent the total mass, the red lines and symbols the stellar mass, while the green ones are for dark matter. We plot the true profiles with solid lines, and use dashed lines with different markers to indicate the model results from mock data with different viewing angles $(\theta,\varphi)$. The blue triangles with error bars in the top of each panel represent the typical error bars ($1\sigma$ confidence levels, see Equation~\ref{1sigma}) of the model estimated total mass. The dots in the bottom of each panel show the relative average difference $1-\langle M_{\rm true}(\textless r)/M_{\rm model}(\textless r)\rangle$ between true values and model estimated values from our mock data with five different projections, and the corresponding error bars represent average confidence levels within $1\sigma$. The vertical dashed line represents $\overline{R_{\rm e}}$ of each galaxy and the value of maximum data coverage $R_{\rm m}$, which means the points outside this value are not reliable, is shown as text. As seen in Figure~\ref{mass}, the estimation of total mass within $\overline{R_{\rm e}}$ is good for all galaxies (average bias $\textless15$ percent) except P2. However the stellar mass is systematically underestimated and the dark matter mass overestimated, especially in the inner parts of galaxies.

In order to quantify the difference between true profiles and model results, we compare in Figure~\ref{mass-all} the enclosed mass within $\overline{R_{\rm e}}$ with the true values of all 45 mock observations. The blue markers represent different viewing angles, from face-on to edge-on are circles, triangles, squares, diamonds, pentagrams. The red triangles with error bars have the same meaning as the error bars in Figure~\ref{mass}. For each galaxy except P2, the estimation of total mass $M_{\rm tot}(\textless \overline{R_{\rm e}})$ is satisfactory, with average relative deviations being inside $\pm15$ percent region. The stellar mass $M_*(\textless \overline{R_{\rm e}})$ is $\sim24$ percent lower than the true value on average and dark matter mass $M_{\rm dark}(\textless \overline{R_{\rm e}})$ is $\sim38$ percent higher.

The potential in our model is generated by both the stellar mass and dark matter mass. The stellar kinematics, in principle, can only constrain the total mass profile $M_{\rm tot}(r)$. The only way to separate stellar mass and dark matter is based on the difference in their density slopes $\gamma$ (see Figure~\ref{mass-density-slope}) . Since we use the NFW profile to construct the dark matter potential, the inner density slopes of our model results in $\gamma$ approximate to $-1$, while the true profiles are more likely to be ``cored'' ($\gamma=0$). Thus we have good recovery for total mass, but the stellar mass and dark matter mass estimates have larger uncertainties due to our choice of dark matter model and to the degeneracy between them. By changing the NFW dark matter profile to the generalized NFW (gNFW, \citealp{Cappellari2013}), whose inner density slopes are not fixed, we re-run 9 mock data sets (one for each galaxy). The average underestimation of stellar mass decrease from $\sim24$ to $\sim13$ percent and the average overestimation of dark matter goes down from $\sim38$ to $\sim18$ percent.

In Figure~\ref{mass-all}, we show comparisons between our Schwarzschild modelling and an oblate JAM technique \citep{Li2016}. We use the same mock data as input to the two different methods. The green and dark symbols represent JAM results, corresponding to the red and blue markers for the results from Schwarzschild modelling. The recovery of total mass $M_{\rm tot}(\textless \overline{R_{\rm e}})$ is also good for JAM modelling. The scatter of relative deviations of all 45 mock data sets are $18$ percent for Schwarzschild modelling and $16$ percent for JAM modelling, which means these two different methods have similar abilities to recover the total mass. However, JAM modelling gives very small error bars, which seems to be optimistic. JAM modelling can yield more narrowly constrained parameters purely because of more restrictive and sometimes ad-hoc assumptions, for example on the velocity anisotropy which is left completely free in Schwarzschild modelling. For the stellar mass and dark matter mass, both techniques give large uncertainties. Using a gNFW profile for dark matter, JAM modelling seems to underestimate dark matter mass and overestimate stellar mass slightly.

Our mass accuracy is consistent with \citet{Thomas2007}. They construct axisymmetric Schwarzschild models for N-body merger remnants using a different orbit sampling method \citep{Thomas2004}. They find the total mass has a bias of between 3 to 20 percent for edge-on views and up to 50 percent for face-on views around $1\sim2R_{\rm e}$. They also adopt the NFW profile for dark matter and find a systematic underestimation for stellar mass: $(M_*/L)_{\rm model}/(M_*/L)_{\rm true}$ is between 0.5 and 0.9.
\subsection{Morphology}
\label{sec6.2}

\begin{figure*}
\begin{center}
	\includegraphics[width=16cm]{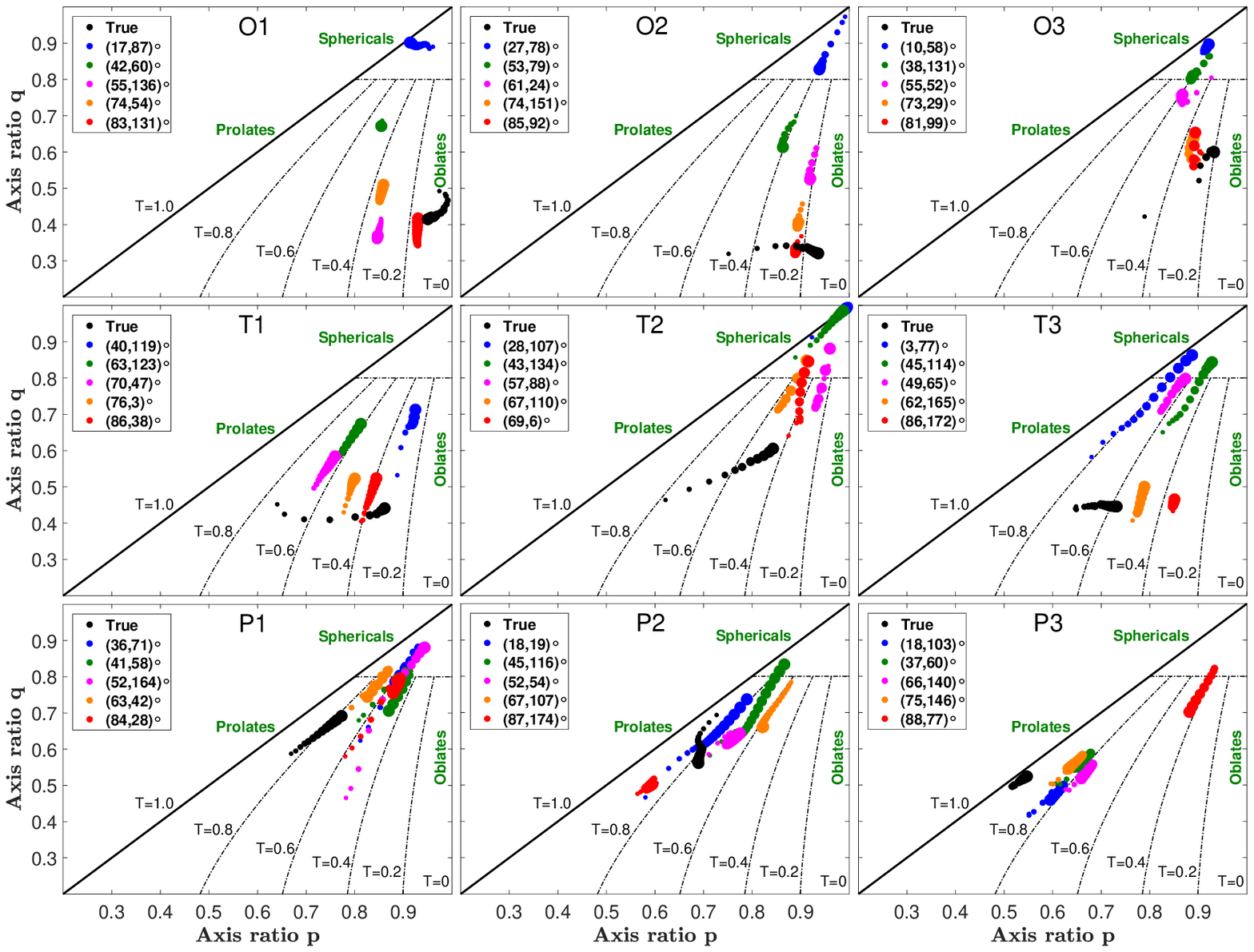}
    \caption{The variation of intrinsic shapes with radius for each galaxy. Within the range $0.5\overline{R_{\rm e}}$ to $2\overline{R_{\rm e}}$, we plot for each galaxy $p=b/a$ versus $q=c/a$ measured at different radii, with larger symbol sizes indicating larger radii. Top panels: oblate galaxies O1, O2 and O3. Middle panels: triaxial galaxies T1, T2 and T3. Bottom panels: prolate galaxies P1, P2 and P3. The black dots are for true shapes, while the blue, magenta, green, orange, red dots represent the shapes recovered from mock data of face-on to edge-on views $(\theta,\varphi)$. The dashed curves and the horizontal dashed lines divide the figures into different regions: $T=0\sim0.2$, $T=0.2\sim0.4$, $T=0.4\sim0.6$, $T=0.6\sim0.8$, $T=0.8\sim1.0$ and nearly spherical regions.}
    \label{shape}
\end{center}
\end{figure*}

\begin{figure*}
\begin{center}
	\includegraphics[width=16cm]{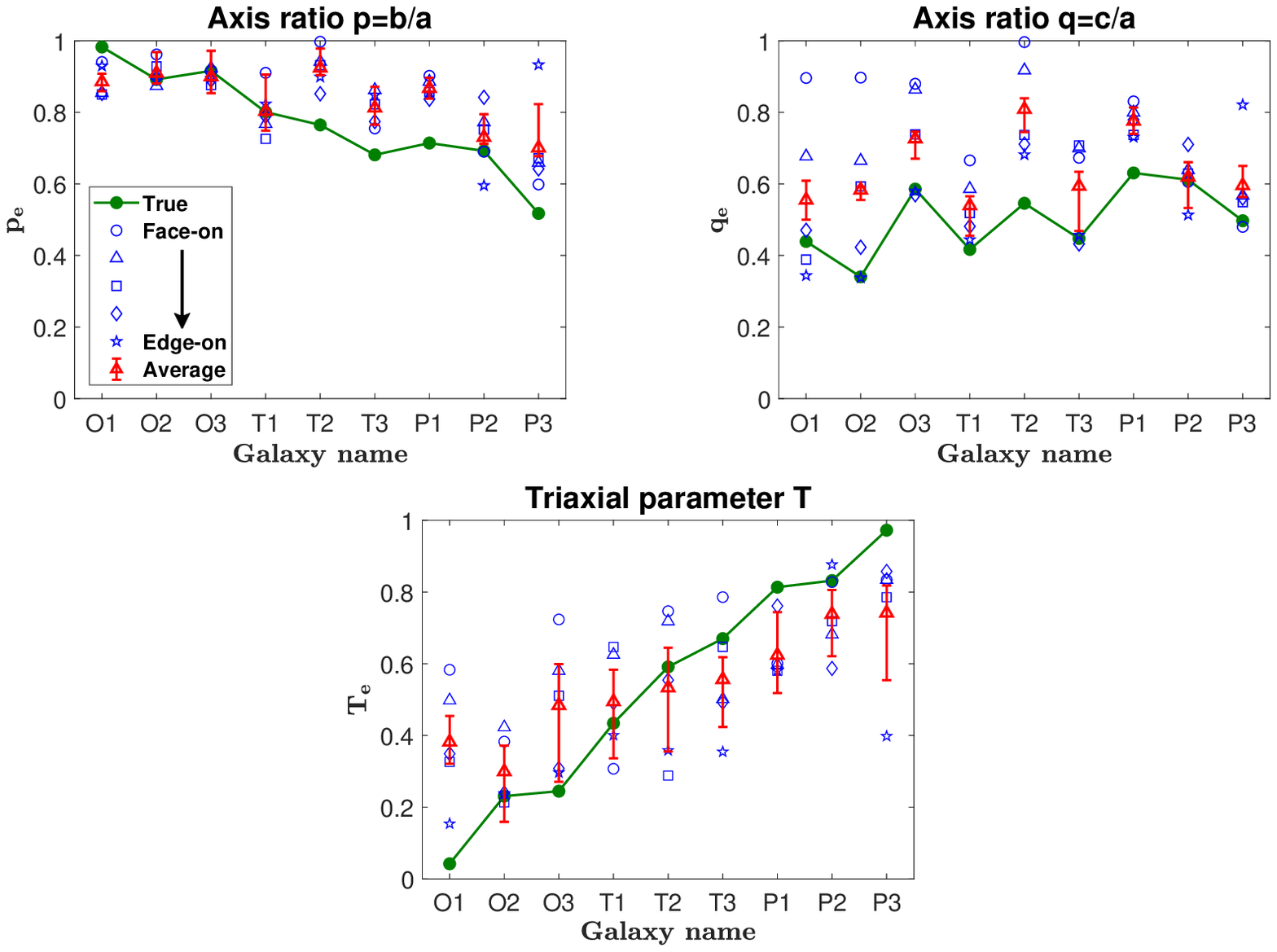}
    \caption{The comparison of intrinsic shapes and triaxial parameter at $\overline{R_{\rm e}}$ between model results and true values. Top left: axis ratio $p_{\rm e}$. Top right: axis ratio $q_{\rm e}$. Bottom: Triaxial parameter $T_{\rm e}$. The horizontal axes mean galaxy names, from left to right is: oblate galaxies O1, O2 and O3; triaxial galaxies T1, T2 and T3; prolate galaxies P1, P2 and P3. The green dots represent the true values calculated from simulation directly, and the circles, triangles, squares, diamonds and pentagrams blue markers are for those recovered by mock data from face-on to edge-on views. The red triangles with error bars show the mean values of results from mock data with five different projections and average confidence levels within $1\sigma$ .}
    \label{shape-all}
\end{center}
\end{figure*}

The different intrinsic shapes of the nine simulated galaxies make them a good sample with enough variations to assess estimation of triaxiality. Here we use the axis ratios $p=b/a$, $q=c/a$ and corresponding triaxial parameter $T=(1-p^2)/(1-q^2)$ to describe galaxy morphology, where $a$, $b$ and $c$ are the major, intermediate and minor axes of the three-dimensional luminosity distribution of the galaxy. Since the stellar potential is generated by a superposition of three-dimensional Gaussians (see $\S$~\ref{sec4.1}), the equipotential planes are usually not regular ellipsoids. Since we choose to ignore any misalignment between different Gaussians, we can simply calculate the model's intrinsic shapes at different radii by applying the least-square ellipse fitting method to luminous contours on projected planes. We thus obtain $p$ as a function of radius $r$, $p(r)$, from the $x$-$y$ plane and obtain $q(r)$ from the $x$-$z$ plane.

The variation of intrinsic shapes with radius for all nine galaxies is shown in Figure~\ref{shape}. We show variations of intrinsic shapes from $0.5\overline{R_{\rm e}}$ to $2\overline{R_{\rm e}}$. The black dots are for true shapes, while the blue, magenta, green, orange, red dots represent the shapes recovered from mock data of face-on to edge-on views. The size of the dots indicate the radius: larger symbols represent larger radii. For oblate and triaxial galaxies, the model estimates improve with increasing inclination $\theta$. The model shapes constrained by edge-on mock data (red) are close to the real morphology (black). This is because we know the real kinematic information well if we observe galaxies dominated by minor-axis rotation in edge-on views, and the modelling thus gives a better restriction for the axis ratio $q$ (see Equation~\ref{transformation}). For the prolate galaxy P3, we find that the model estimate with an edge-on view is quite different. For the viewing angles $(\theta,\varphi)=(88,77)^{\circ}$, we are nearly observing this galaxy along the major-axis. We therefore are unable to obtain enough information about the major-axis rotation, which dominates prolate galaxies. For all the galaxies, we find that some models tend to overestimate $p$ and $q$, which means they are slightly biased towards sphericals.

We show a quantitative comparison of the true and model values for intrinsic shapes and the triaxial parameter at $\overline{R_{\rm e}}$  in Figure~\ref{shape-all}. The green dots represent the true values calculated from simulation directly, and the circles, triangles, squares, diamonds and pentagrams blue markers are for those estimated from mock data face-on to edge-on views. The red triangles with error bars show the mean values of results from mock data with five different projections and average confidence levels within $1\sigma$. For the axis ratio $p$, four of nine galaxies (O2, O3, T1, P2) give relatively good recoveries, while four galaxies overestimate $p$ by $0.13 \sim 0.18$ and one galaxy underestimate $p$ by $0.10$. For the axis ratio $q$, only the galaxy P2 shows a good recovery, while other galaxies have average overestimations between $0.10$ and $0.26$. The mean bias of these two parameters are $\Delta p=0.07$ and $\Delta q=0.14$. Although there are some systematic biases of $p$ and $q$, we can still see a trend in the distribution of triaxial parameter $T$: the model recovered $T_{\rm e}$ rises as the true $T_{\rm e}$ rises but with a shallower slope. For galaxies O1, O2, O3 and T1, $T_{\rm e}$ is overestimated, while it is underestimated for T2, T3, P1, P2, P3. This means that on average the modelling tends to make galaxies more triaxial.

\subsection{Circularity}
\label{sec6.3}
\begin{figure*}
\begin{centering}
	\includegraphics[width=16cm]{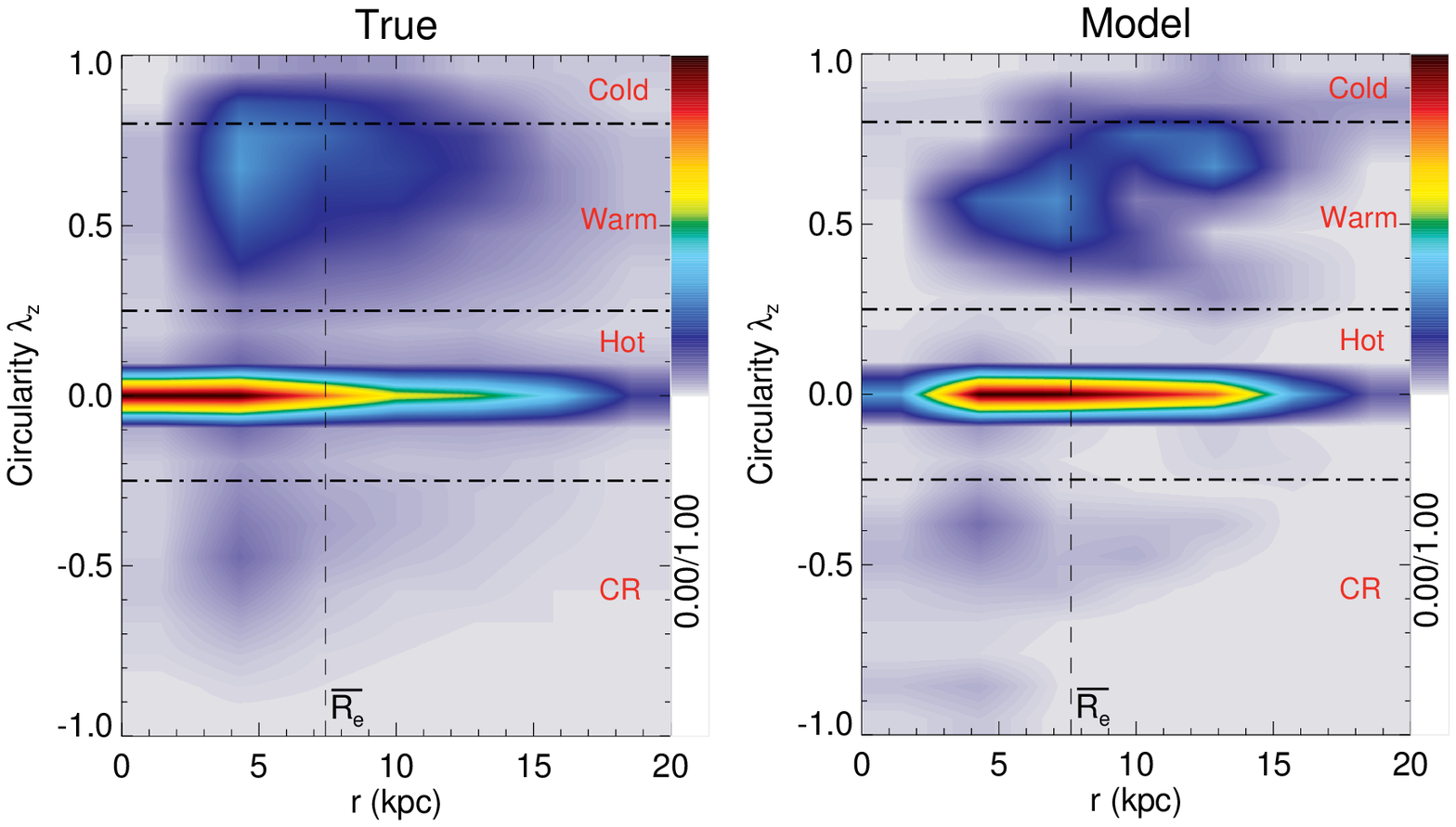}
    \caption{The comparison between true circularity distribution on the 2D coordinate space $f(r,\lambda_z)$ of an example galaxy T1 (left) and the corresponding model estimate from the mock data set with viewing angles $(70,47)^{\circ}$ (right). Each map is equally divided into $n_r\times n_{\rm \lambda}=10\times21$ rectangle bins. Each colored pixel represents the probability density of orbits in the bins, from blue (low density) to red (high density). The maps are smoothed and the color bar is linear. The horizontal dashed lines divide the maps into cold orbits ($\lambda_z\ge0.8$), warm orbits ($0.25<\lambda_z<0.8$), hot orbits ($-0.25\le\lambda_z\le0.25$) and counter-rotating orbits ($\lambda_z<-0.25$), while the vertical dashed lines indicate $\overline{R_{\rm e}}$.}
    \label{lambdaz-2d}
\end{centering}
\end{figure*}
\begin{figure*}
\begin{centering}
	\includegraphics[width=16cm]{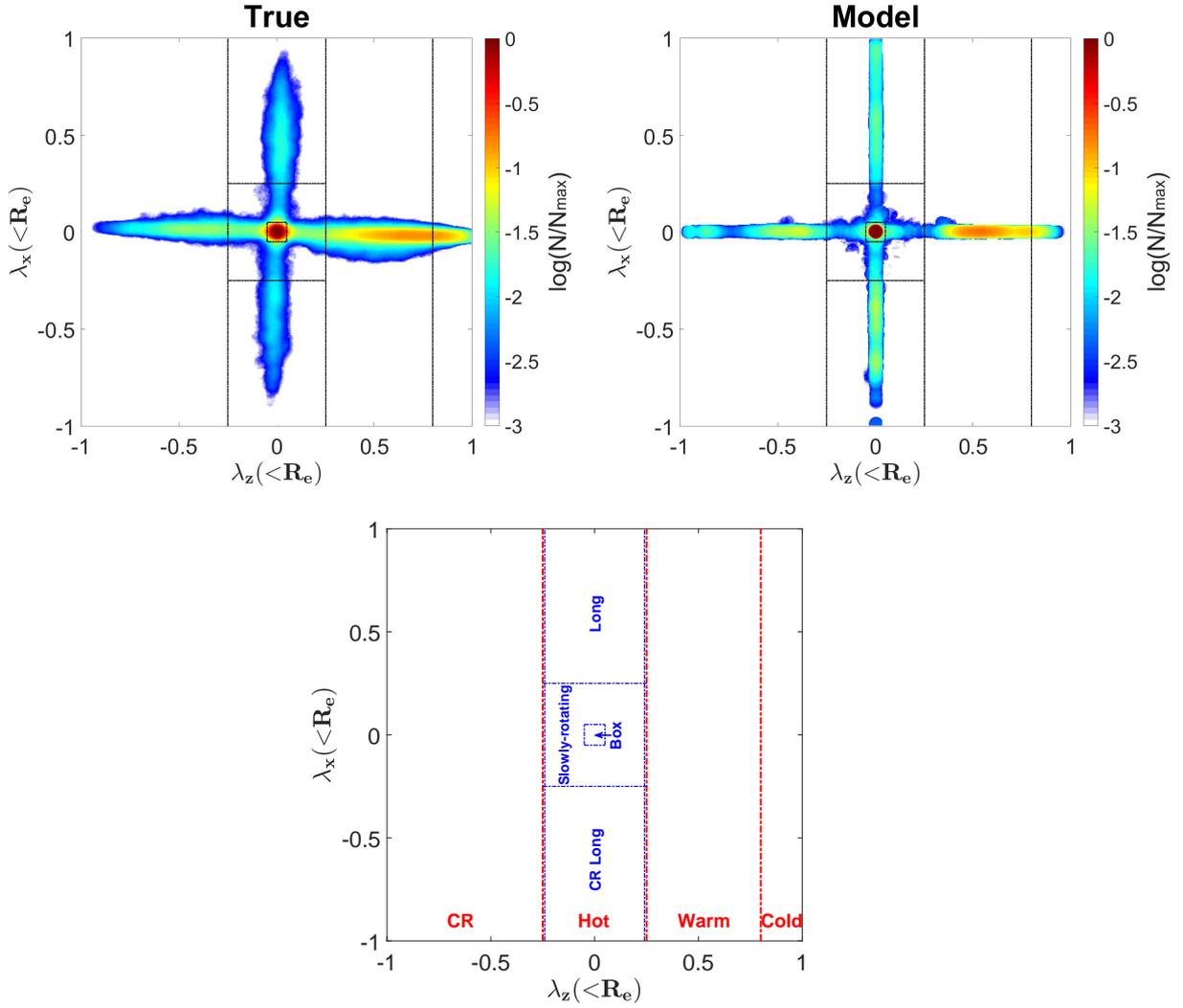}
    \caption{The comparison between true circularity distribution within $\overline{R_{\rm e}}$ in different axes $\lambda_x(\textless \overline{R_{\rm e}})$ versus $\lambda_z(\textless \overline{R_{\rm e}})$ of an example galaxy T1 (top left) and the corresponding model estimate from the mock data set with viewing angles $(70,47)^{\circ}$ (top right). The orbit probability density $\log(N/N_{\rm max})$ is indicated by the color bar. We plot the dashed lines in the bottom panel as the same as in top panels to show how we divide orbits. The red lines means we divide orbits with different $\lambda_z$ to be cold ($\lambda_z\ge0.8$), warm ($0.25<\lambda_z<0.8$), hot ($-0.25\le\lambda_z\le0.25$) and counter-rotating components ($\lambda_z<-0.25$), while the blue lines mean we separate hot orbits to be four different components as box orbits and long-axis tubes can not be separated from $\lambda_x$ : prograde long-axis tubes ($\lambda_x>0.25,\left|\lambda_z\right|\le0.25$), counter-rotating long-axis tubes ($\lambda_x<-0.25,\left|\lambda_z\right|\le0.25$), box orbits ($\left|\lambda_x\right|\le0.05,\left|\lambda_z\right|\le0.05$) and slowly-rotating orbits ($\left|\lambda_x\right|,\left|\lambda_z\right|\le0.25,\left|\lambda_x\right|$ or $\left|\lambda_z\right|>0.05$).}
    \label{lambdaz-vs-x}
\end{centering}
\end{figure*}
\begin{figure*}
\begin{centering}
	\includegraphics[width=14cm]{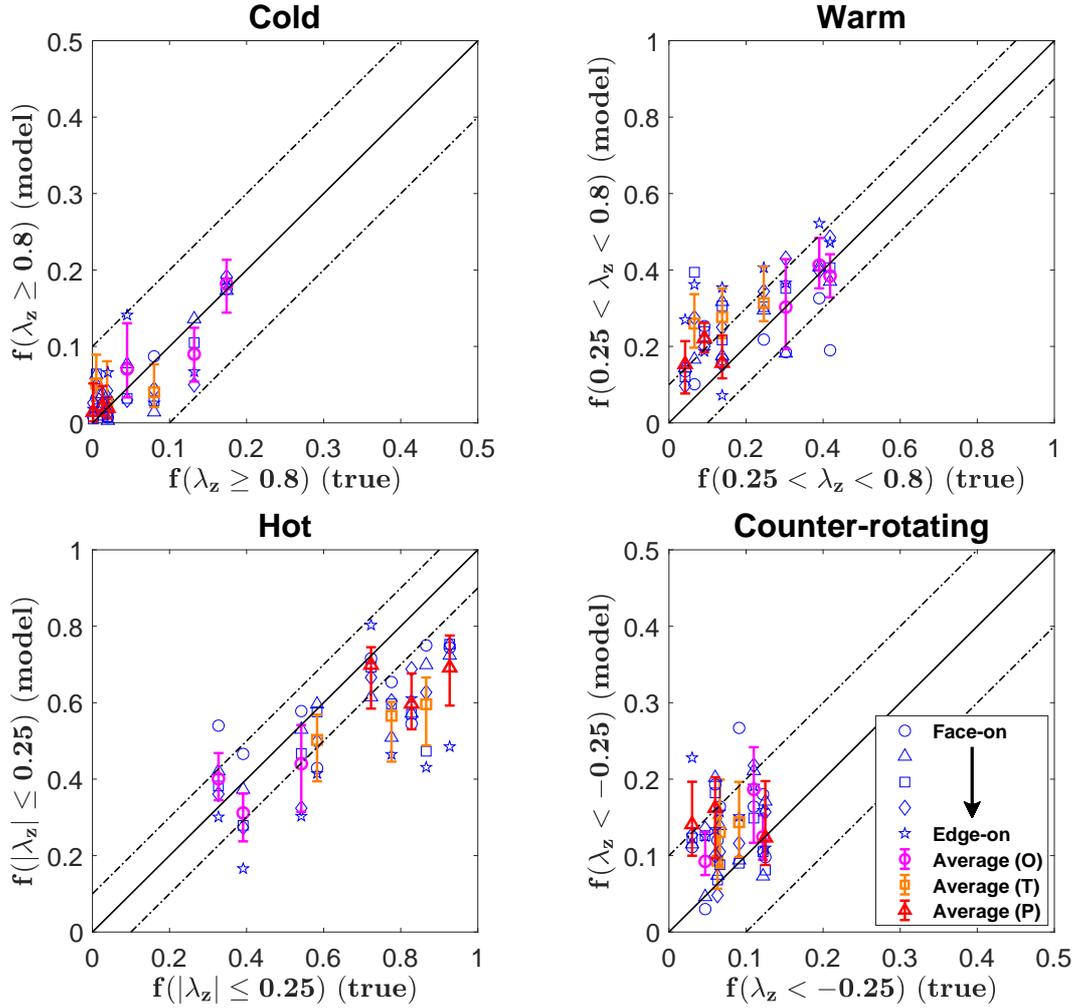}
    \caption{One-to-one comparison of the true and model estimated luminosity fractions of cold, warm, hot, and counter-rotating orbits within $\overline{R_{\rm e}}$. Top left: cold components ($\lambda_z\ge0.8$). Top right: warm components ($0.25<\lambda_z<0.8$). Bottom left: hot components ($-0.25\le\lambda_z\le0.25$). Bottom right: counter-rotating components ($\lambda_z<-0.25$). The circles, triangles, squares, diamonds and pentagrams blue markers represent model results with face-on to edge-on views. The magenta circles, orange squares and red triangles with error bars show the mean values of results from mock data with five different projections and average confidence levels within $1\sigma$ for oblates, triaxials and prolates. The solid lines represent equal values, while the dashed lines are $\pm 0.1$ away from the solid lines. Note that the axes ranges are different in four panels.}
    \label{lambdaz-all}
\end{centering}
\end{figure*}

\begin{figure*}
\begin{centering}
	\includegraphics[width=14cm]{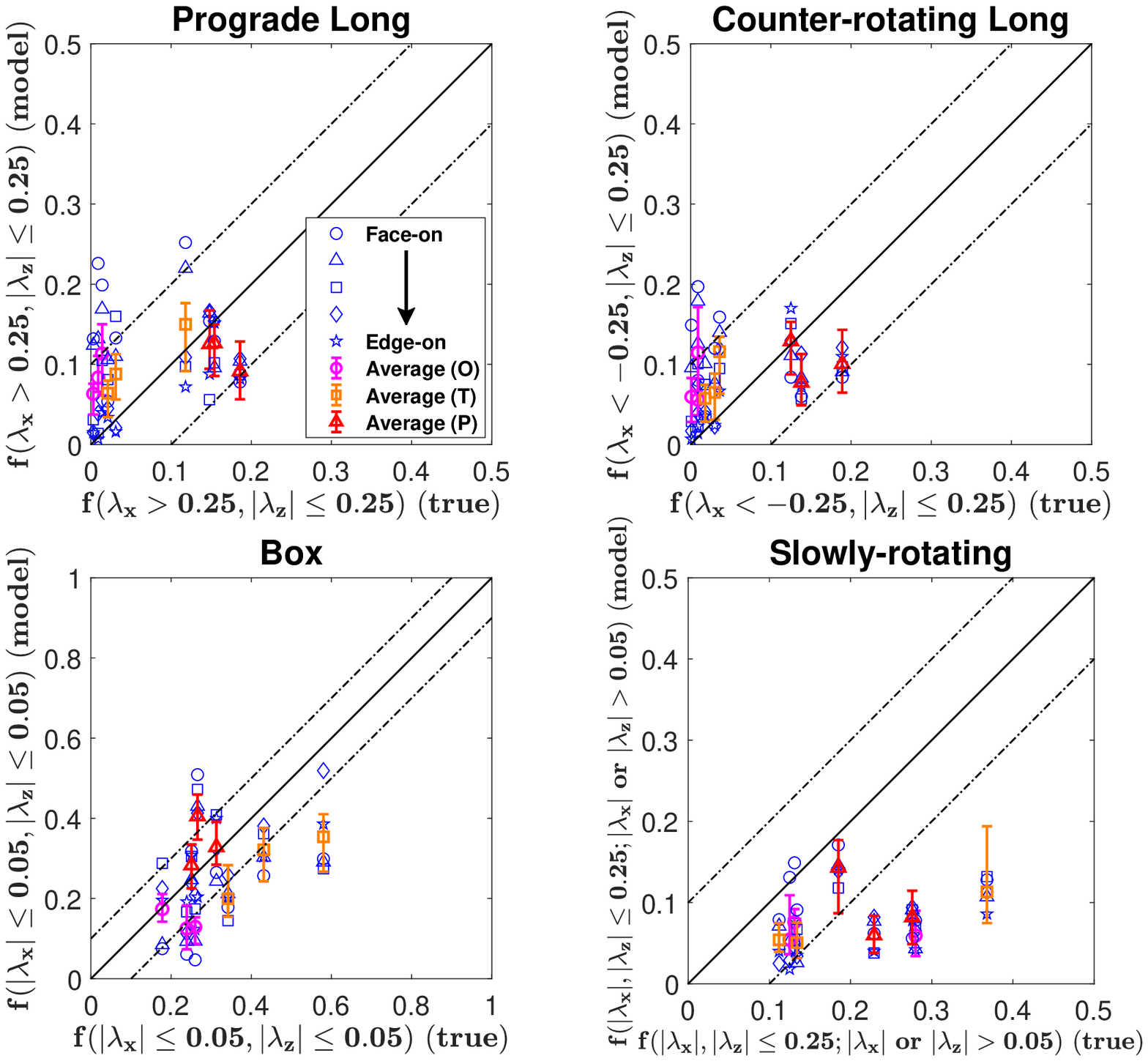}
    \caption{One-to-one comparison of the true and model estimated luminosity fractions of long axis tubes, box orbits and slowly-rotating orbits within $\overline{R_{\rm e}}$. Top left: prograde long-axis tube components ($\lambda_x>0.25,\left|\lambda_z\right|\le0.25$). Top right: counter-rotating long-axis tube components ($\lambda_x<-0.25,\left|\lambda_z\right|\le0.25$). Bottom left: box components ($\left|\lambda_x\right|\le0.05,\left|\lambda_z\right|\le0.05$). Bottom right: slowly-rotating components ($\left|\lambda_x\right|,\left|\lambda_z\right|\le0.25,\left|\lambda_x\right|$ or $\left|\lambda_z\right|>0.05$). The symbols and lines have the same meaning as those in Figure~\ref{lambdaz-all}.}
    \label{lambdax-all}
\end{centering}
\end{figure*}

As mentioned in $\S$~\ref{sec4.5}, the $\lambda_z$ circularity parameter is used to characterize the orbits. In the modelling, we calculate the circularity parameter for each orbit. Based on the orbit weights from the best-fitting model, we obtain the probability density distribution on the 2D coordinate space $f(r,\lambda_z)$. In the simulations, the orbital circularity of each stellar particle is obtained through running 20-orbit revolutions and recording time-averaged quantities of $L_z$, $r$ and $V_{\rm rms}$. In order to do so, we freeze the gravitational potential of the galaxy at the current snapshot and use the phase space information of given stellar particles as their initial condition. A second-order leapfrog integrator is adopted to update the positions and velocities during orbital evolution \citep{Dehnen2011}. Thus we can also obtain the distribution $f(r,\lambda_z)$ from the simulations.

We show the comparison of this distribution for an example galaxy T1 in Figure~\ref{lambdaz-2d}. The left panel shows the true circularity distribution while the right panel shows the model estimate for the mock data set with viewing angles $(70,47)^{\circ}$. Each map is equally divided into $n_r\times n_{\rm \lambda}=10\times21$ rectangle bins. Each colored pixel represents the probability density of orbits in the bins, from blue (low density) to red (high density). The maps are smoothed and the color bar is linear. In order to further quantify the model estimate, we divide orbits into different components based on $f(r,\lambda_z)$. Following \citet{Zhu2018b}, we classify orbits with different $\lambda_z$ to be: dynamically cold ($\lambda_z\ge0.8$), dynamically warm ($0.25<\lambda_z<0.8$), dynamically hot ($-0.25\le\lambda_z\le0.25$) and counter-rotating ($\lambda_z<-0.25$) components. From our models, the hot orbits mainly consist of box orbits and long-axis tube orbits, while the warm and cold orbits are short-axis tube orbits. These four different components are indicated by the horizontal dashed lines in the maps. We can see the model orbit distribution $f(r,\lambda_z)$ is a reasonable match to the true distribution.

In Figure~\ref{lambdaz-2d}, there is a clear peak around $\lambda_z\sim 0$, which comprises mainly box orbits and long-axis tube orbits that can not be separated in the distribution of $\lambda_z$. For the purpose of distinguishing them, we also need to examine the distribution of $\lambda_x$, which quantifies rotation about the major axis. We calculate both $\lambda_z$ and $\lambda_x$ for all active orbits within $\overline{R_{\rm e}}$ for the same example galaxy and plot the probability density distributions in Figure~\ref{lambdaz-vs-x}. The top left panel shows the true distribution while the top right panel gives the model estimate. The bottom panel indicate how we separate orbits with different circularities. In addition to separating cold, warm, hot and counter-rotating components based on $\lambda_z$, we also separate the hot orbits based on $\lambda_x$: prograde long-axis tube components ($\lambda_x>0.25$), counter-rotating long-axis tube components ($\lambda_x<-0.25$), box components ($\left|\lambda_x\right|\le0.25$) and slowly-rotating components ($\left|\lambda_x\right|,\left|\lambda_z\right|\le0.25,\left|\lambda_x\right|$ or $\left|\lambda_z\right|>0.05$).

We show a comparison of cold, warm, hot, and counter-rotating orbit fractions within $\overline{R_{\rm e}}$ in Figure~\ref{lambdaz-all}. The magenta circles, orange squares and red triangles with error bars represent the mean values and $1\sigma$ uncertainties for oblates, triaxials and prolates. The blue symbols with different markers represent model values obtained from mock data with different viewing angles $(\theta,\varphi)$. The solid lines mean equal values, while the dashed lines are $\pm 0.1$ away from the solid lines. Note that the coordinate ranges are different in each panel. For $\lambda_z$, the comparison of cold components is quite good, with an average bias $\approx 0$. The warm fractions are $0.07$ overestimated and the counter-rotating fractions are $0.05$ overestimated on average, while the hot fractions have an average underestimation equal to $0.12$. Ignoring galaxy morphologies, these results are consistent with \citet{Zhu2018b}, whose sample mainly consist of spiral galaxies, together with some oblate and triaxial early-type galaxies.

Comparison of the luminosity fractions of long-axis tubes, box orbits and slowly-rotating orbits is separately shown in Figure~\ref{lambdax-all} (symbols are the same as in Figure~\ref{lambdaz-all}). For oblates and triaxials with face-on views, modelling tends to overestimate both prograde and counter-rotating long-axis tubes, while correspondingly box orbits are underestimated. For prolate galaxies, the long-axis tube fractions for most viewing angles are lower than the true values, while the box orbit fraction is higher. These results indicate a clear modelling degeneracy between long-axis tube orbits and box orbits. Slowly rotating orbit fractions are underestimated for almost all mock data sets, and is the main reason behind the bias of hot orbit fractions.

\subsection{Velocity anisotropy and tangential fraction}
\label{sec6.4}

\begin{figure*}
\begin{centering}
	\includegraphics[width=16cm]{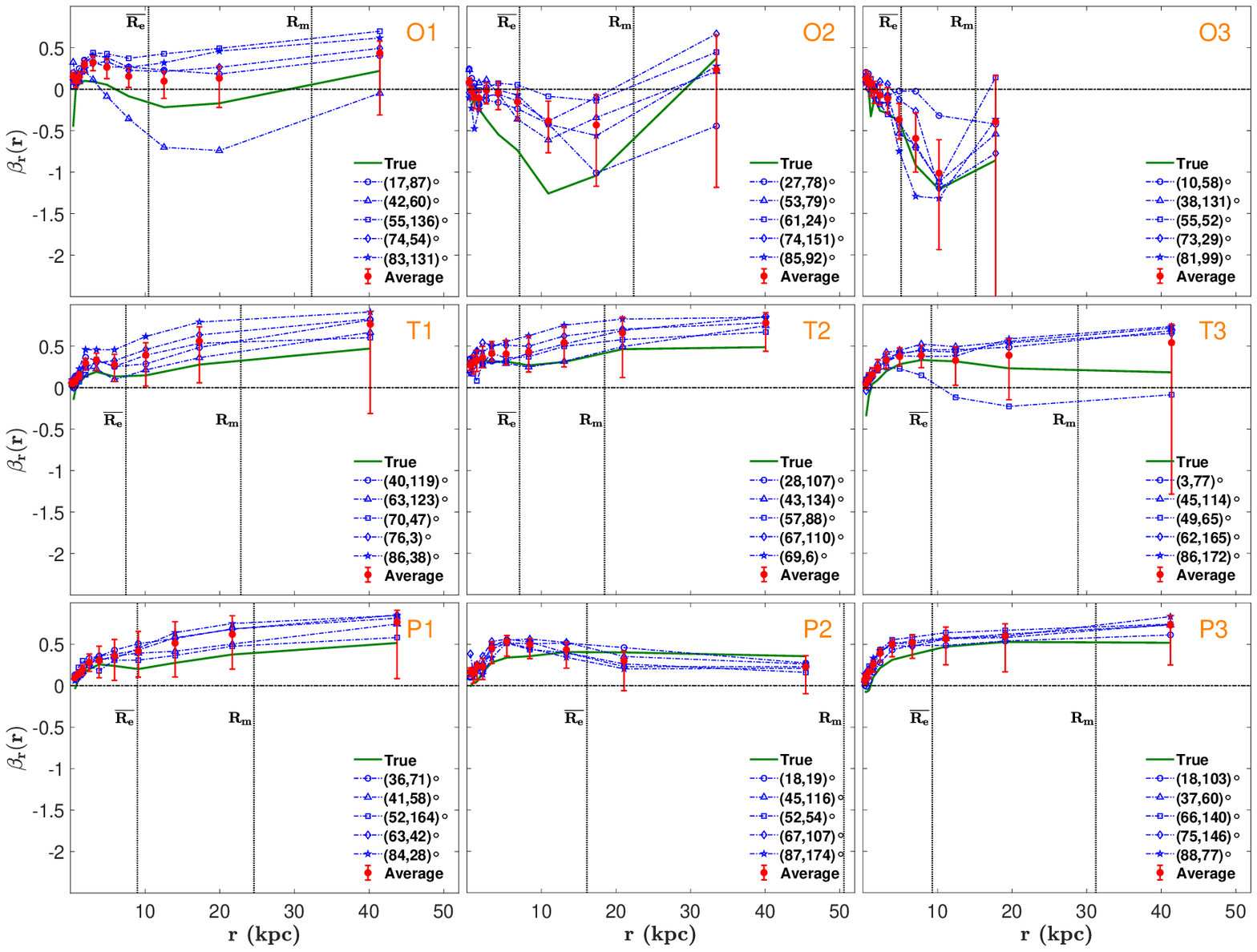}
    \caption{The comparison of velocity anisotropy profiles $\beta_r(r)$ between true values and those obtained by our models. Top panels: oblate galaxies O1, O2 and O3. Middle panels: triaxial galaxies T1, T2 and T3. Bottom panels: prolate galaxies P1, P2 and P3. The green lines represent the true velocity anisotropy profiles and blue dashed lines with different markers are for model recoveries obtained from mock data with different viewing angles $(\theta,\varphi)$. The red dots with error bars show the mean values of results from mock data with five different projections and average confidence levels within $1\sigma$. The horizontal dashed lines are isotropic lines ($\beta_r=0$), separate tangential regions ($\beta_r<0$) and radial regions ($\beta_r>0$). The vertical dotted lines indicate $\overline{R_{\rm e}}$ and maximum data coverage $R_{\rm m}$.}
    \label{betar}
\end{centering}
\end{figure*}

\begin{figure*}
\begin{centering}
	\includegraphics[width=16cm]{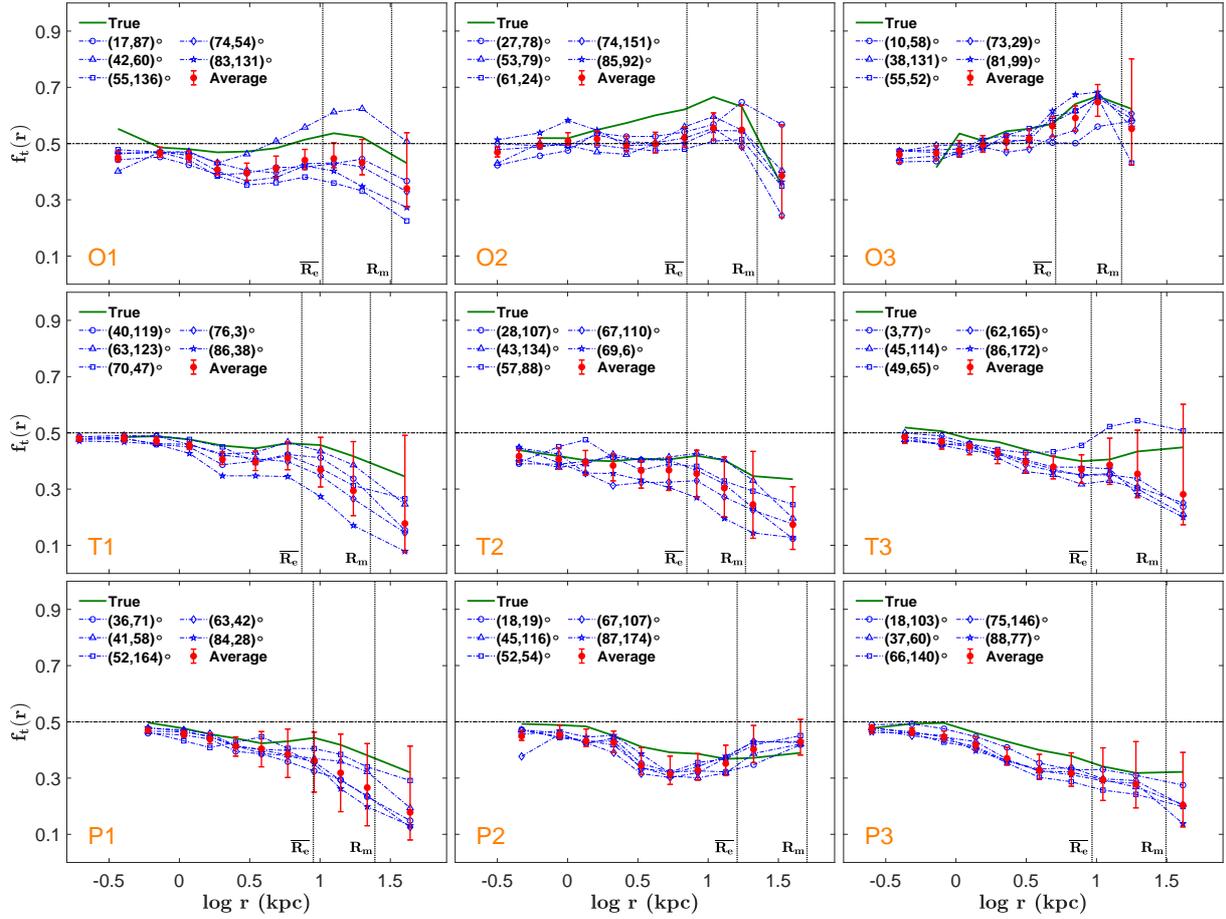}
    \caption{The comparison of tangential fraction profiles $f_t(r)$ versus $\log\ r$ between true values and those obtained by our models. Top panels: oblate galaxies O1, O2 and O3. Middle panels: triaxial galaxies T1, T2 and T3. Bottom panels: prolate galaxies P1, P2 and P3. The symbols and lines have the same meaning as those in Figure~\ref{betar}. Due to the limited resolution of the Illustris simulation, the green lines do not extend to the most inner regions for galaxy O2, O3 and T1.}
    \label{tfrac}
\end{centering}
\end{figure*}

\begin{figure*}
\begin{centering}
	\includegraphics[width=9cm]{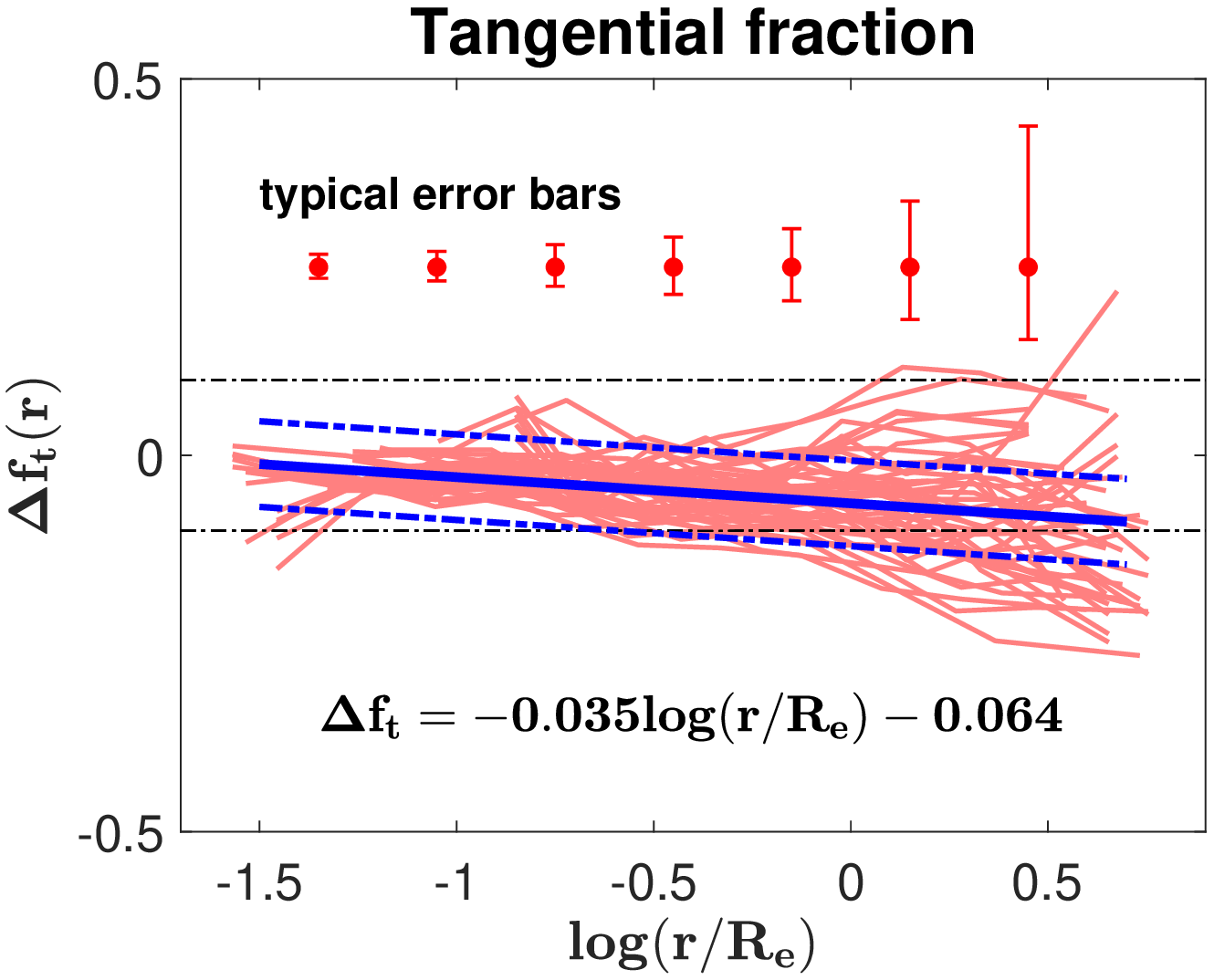}
    \caption{The error of tangential fraction profile $\Delta f_t(r)=f_t(r)_{\rm model}-f_t(r)_{\rm true}$ between model results and true values versus $r/R_{\rm e}$. Each red fold line means $f_t(r)$ obtained from each mock data set. The blue solid line are the fitted line of all models, corresponding to the equation in the figure, while the blue dashed lines represent the errors in linear fitting program. The black dashed lines represent where $\Delta f_t(r)=\pm 0.1$. The red dots with error bars represent the typical error bars of model estimated tangential fractions at different radii.}
    \label{tfrac-all}
\end{centering}
\end{figure*}

We calculate the luminosity-weighted velocity anisotropy $\beta_r(r)$ along spherical radius $r$ in the same way from the simulations and from our models. The three-dimensional space is divided into different cells. For the cell $i$ located at radius $r_i$, we can obtain the stellar mass $M_*(r_i)$ and the velocity anisotropy $\beta_r(r_i)$. Thus we have $\beta_r(r)=\Sigma_i[M_*(r_i)\times \beta_r(r_i)]/M_*(r)$

In Figure~\ref{betar}, we compare $\beta_r(r)$ from the simulated galaxies with $\beta_r(r)$ from the models. The green lines represent the true velocity anisotropy profiles and blue dashed lines with different markers are model values obtained from the mock data at different viewing angles. The red dots with error bars show the mean values of results from mock data with the five different projections and the average confidence levels within $1\sigma$. Due to the limited resolution of the Illustris simulation, the green lines do not extend to the most inner regions ($\sim0.5$ kpc) of some galaxies. The oblate galaxies tend to be tangential at $\sim10$ kpc, indicating that they are dominated by short-axis tube orbits in these regions. Triaxials and prolates become more and more radial with increasing radius. Our models roughly match the true $\beta_r(r)$ profiles for our different types of galaxies, although not without some bias.

In order to analyse the model biases, we plot the tangential fraction $f_t$ (see definition in Equation~\ref{tangential}) in Figure~\ref{tfrac}. We change the linear radius to log radius in this figure. The symbols are the same as Figure~\ref{betar}. The model profiles track the true profiles, but with some underestimation for most galaxies, especially in the outer parts. This means that our models are generally more radial than our simulated galaxies. \citet{Long2012} also find a similar bias towards radial anisotropy, using a different dynamical method Made-to-Measure (M2M).

We calculate the differences in tangential fraction $\Delta f_t(r)=f_t(r)_{\rm model}-f_t(r)_{\rm true}$ between the model and true profiles and show them in Figure~\ref{tfrac-all}. By performing a simple straight line fit for $\Delta f_t(r)$, we find a clear systematic trend. The tangential fraction does not have obvious bias in the inner regions. However, with increasing radius, the underestimation becomes larger. The average bias $\Delta f_t(r)$ at $\overline{R_{\rm e}}$ is around 0.064 and reaches 0.1 at large radii.
\subsection{Testing the effect of initial conditions}
\label{sec6.5}
How to select orbits is always important in Schwarzschild modelling. We do not attempt to vary the method by which the orbit initial conditions are created, but examine varying the number of orbits and assessing the impact.

Our results in $\S$~\ref{sec6.1} to $\S$~\ref{sec6.4} are obtained from empirical settings of the initial conditions: $n_E \times n_{I_2} \times n_{I_3}=21\times10\times7$. In order to check whether some systematic biases are caused by the settings, we performed some tests on the effect of orbit sampling. We select three observations: galaxy O1 with viewing angles $(\theta,\varphi)=(42,60)^{\circ}$, galaxy T2 with viewing angles $(\theta,\varphi)=(69,6)^{\circ}$ and galaxy P1 with viewing angles $(\theta,\varphi)=(84,28)^{\circ}$. We change the values of $n_E\times n_{I_2} \times n_{I_3}$ to be $15\times7\times5$ (fewer input orbits) and $31\times15\times10$ (more input orbits), and then rerun the models. We check the estimates of mass distributions, morphologies and orbit distributions for these new models, and do not find any significant differences. The only noticeable difference is that the $1\sigma$ confidence level given by the modelling becomes larger with increasing numbers of input orbits. This could be caused by the NNLS program --- the numbers of active orbits (output orbits with nonzero weights) is not directly proportional to the number of input orbits. The estimates do not have significant changes, but more freedom is allowed in the modelling with more orbits sampled. The detailed results concerning active orbits are shown in Table~\ref{active orbits}.

\begin{table*}
\caption{The information of active orbits. From left to right are: (1) the mock data, contain galaxy names and viewing angles; (2) initial conditions $n_E\times n_{I_2} \times n_{I_3}$; (3) the number of input orbits before dithering; (4) the number of active orbits; (5) the fraction of active orbits.}
    \centering
    \begin{tabular}{|c|c|c|c|c|}
    \hline
    Mock data & Initial conditions $n_E\times n_{I_2} \times n_{I_3}$ & Input orbits & Active orbits & Active fraction \\
    \hline
    \multirow{3}*{galaxy O1 with $(42,60)^{\circ}$} & $15\times7\times5$ & 1575 & 332 & $21.1\%$ \\
    ~                                             & $21\times10\times7$ & 4410 & 479 & $10.9\%$ \\
    ~                                             & $31\times15\times10$ & 13950 & 582 & $4.2\%$ \\
    \hline
    \multirow{3}*{galaxy T2 with $(69,6)^{\circ}$} & $15\times7\times5$ & 1575 & 343 & $21.8\%$ \\
    ~                                             & $21\times10\times7$ & 4410 & 494 & $11.2\%$ \\
    ~                                             & $31\times15\times10$ & 13950 & 567 & $4.1\%$ \\
    \hline
    \multirow{3}*{galaxy P1 with $(84,28)^{\circ}$} & $15\times7\times5$ & 1575 & 339 & $21.5\%$ \\
    ~                                             & $21\times10\times7$ & 4410 & 541 & $12.3\%$ \\
    ~                                             & $31\times15\times10$ & 13950 & 696 & $5.0\%$ \\
    \hline
    \end{tabular}
    \label{active orbits}
\end{table*}

\section{Discussion}
\label{sec7}
From the results shown in $\S$~\ref{sec6.1}, Schwarzschild modelling and JAM modelling have similar abilities in being able to recover total mass. Schwarzschild modelling has more freedom, and usually gives a number of reasonable models around the best-fitting model, while JAM modelling tends to underestimate errors due to its strong assumptions. For stellar mass and dark matter mass, both methods give results with large biases. The computer time required in using each scheme is quite different. As Schwarzschild modelling is an orbit-superposition method, it is time-consuming to construct the orbits particularly when dithering is used. For each combination of free parameters, we need to integrate $4410\times5^3\approx5\times10^5$ orbits. A typical time scale for arriving at a best-fitting triaxial Schwarzschild model for our mock galaxies is $2\sim3$ days on a 256 cores cluster, while JAM is around 100 times faster. For investigators who only want to determine the masses of galaxies, using JAM is more efficient, and little mass information will be lost by comparison with using Schwarzschild's method. The advantage of Schwarzschild's method is orbit information that JAM can not provide.

\begin{figure*}
\begin{center}
	\includegraphics[width=16cm]{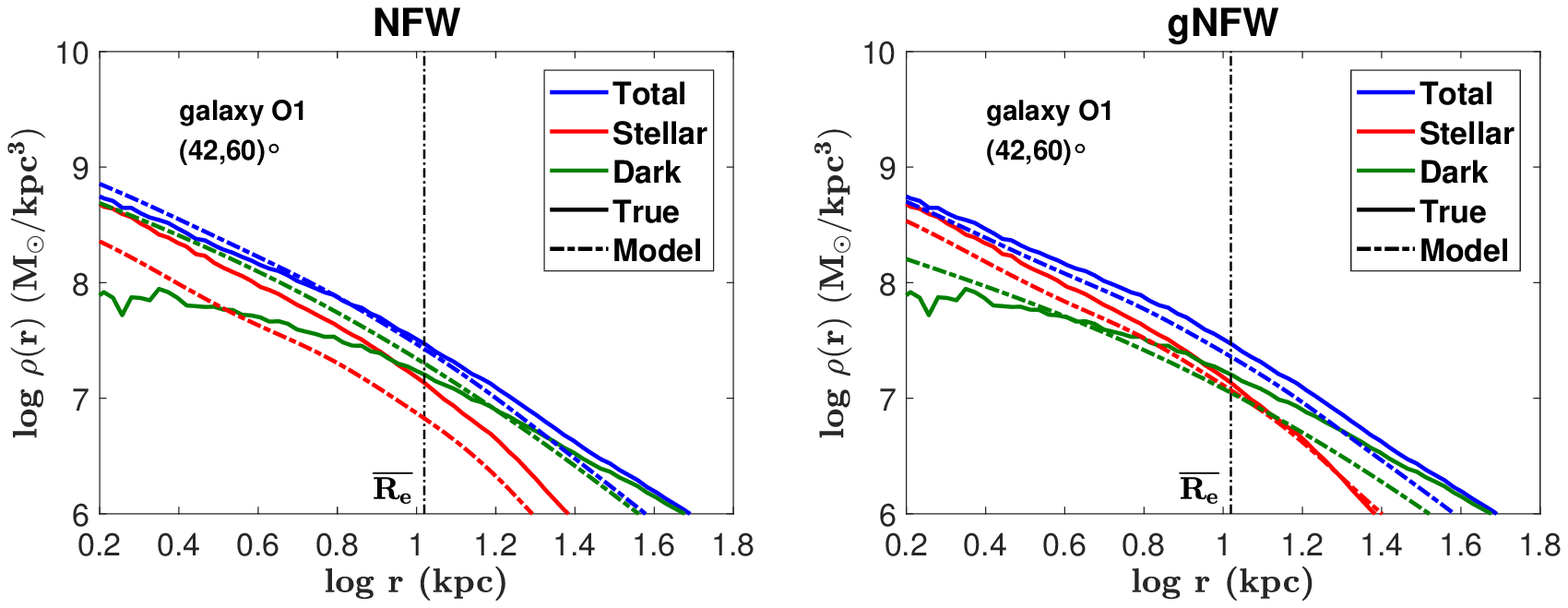}
    \caption{The comparison of the mass density profiles $\rho(r)=\Delta M(r)/(4\pi r^2\Delta r)$ for galaxy O1 with viewing angles $(\theta,\varphi)=(42,60)^{\circ}$ using NFW and gNFW dark matter profiles. The blue lines represent the total mass, the red lines represent the stellar mass, while the green ones are for dark matter. True profiles are indicated by solid lines while the dashed lines show model results. Our best-fitting model with gNFW halo matches the true stellar mass profile and dark matter mass profiles better than the model with NFW halo.}
    \label{NFW-vs-gNFW}
\end{center}
\end{figure*}

Since only the total mass is constrained by the kinematic data, it is difficult to separate the stellar and dark matter components directly from kinematics. The only way to distinguish them is through the inner mass density slope $\gamma$. As our modelling takes the projected surface brightness as input, we are able to give a good estimate of the stellar density slope, but the dark matter density slope can only come from hypothesis. We have assumed that the dark matter mass follows the NFW profile, so the inner dark matter density slope equals $-1$, while the stellar density slope is usually steeper than $-1$. However, the dark matter density slope from simulations is close to 0 (see Figure~\ref{mass-density-slope}). This means that the modelling must control the fraction of stellar mass to ensure that the total mass fits the true value. This directly relates to why the stellar mass has a systematic underestimation. Clearly, a good assumption of the dark matter profile is crucial for estimating these two mass components more accurately. By changing the dark matter profile from NFW to gNFW (generalized NFW), we find the biases in stellar mass and dark matter reduce appreciably. However, there is still a large uncertainty for the inner density slope of dark matter and it is not recovered well. Due to computer time limitations, we only model a few galaxies with the gNFW profile as it has more free parameters. We take galaxy O1 with viewing angles $(42,60)^{\circ}$ as an example to show the difference between density profiles in our modelling by using NFW and gNFW haloes in Figure~\ref{NFW-vs-gNFW}.

In $\S$~\ref{sec6.2}, we find intrinsic shapes have large uncertainties, especially for oblate and triaxial galaxies with face-on viewing angles. We suggest there are two reasons for this. The main reason is that the MGE method can only restrict the minimum value of the axis ratio $q_{\rm min}$ from the projected surface brightness. For face-on views, it is very possible for modelling to overestimate $q$. The other axis ratio $p$ is related to $q$ $(p>q)$, so that model estimates are more nearly spherical. The second reason is that, in our models, dark matter is taken to be spherical while the stellar component is triaxial. This means that the degeneracy between stellar mass and dark matter mass may influence the intrinsic shapes. Overall, we believe that our existing information is not sufficient to constrain the intrinsic shapes very well.

By combining the estimation of intrinsic shapes and circularity $\lambda_x$ distributions, we find the bias of axis ratios have a clear relationship with the degeneracy between long-axis tubes and box orbits: both the overestimation of long-axis tube fractions for face-on views in oblates and triaxials and the underestimation of long-axis tube fractions in prolates requires the intrinsic shapes to become more triaxial, which is consistent with our results in Figure~\ref{shape-all}.

In $\S$~\ref{sec6.3}, there is a systematic bias identified in the hot orbit fractions. We separate hot orbits into four parts and find the bias is mainly caused the slowly-rotating orbits. These orbits act like ``bridges'' that connect box orbits and tube orbits. In galaxy formation, the orbit distribution is usually continuous so the circularity ranges of box orbits and tube orbits could be broader. However, in our modelling, we sample many box orbits on equipotential planes, which means these box orbits will concentrate on the peaks at $\lambda_z=0$ and $\lambda_x=0$. Our models lack these slowly rotating orbits and in that respect our orbit initial conditions are deficient.

We analyse the orbit distributions in $\S$~\ref{sec6.3} and $\S$~\ref{sec6.4}. From the anisotropy parameters $\beta_r$ and $f_t$, we find a systematic trend that the model values are more radial than our simulated galaxies. This trend does not appear in circularity distributions. Although both velocity anisotropy and circularity are orbit properties, what they tell us is different. A key point is that circularity could not distinguish different kinds of box orbits, which are all located on the central peak $\lambda_z\sim 0$ and $\lambda_x\sim 0$. \citet{Vietri1983} studied box orbits in triaxial galaxies and introduced a width parameter to classify box orbits. From their paper, some box orbits are ``wide'', like peanuts; while some are ``narrow'', like pens. These ``narrow'' box orbits could be quite radial, while the ``wide'' box orbits may not be --- they can also have large fractions of tangential velocity components. A possible explanation of our results is that modelling overestimates radial box orbits but underestimate other box orbits. A detailed investigation of the properties of these different orbit families may help us to understand this problem. Since our results are almost independent of sampling orbits (see $\S$~\ref{sec6.5}), the way to reduce the biases may be by finding some useful restriction to the orbit families.

In assessing our work, we are mindful that any results where we are comparing model 6-dimensional data (three positions and three velocities) with our simulated galaxy data are subject to deprojection effects because of the 3-dimensional data (two positions and one velocity) we use as constraints. This means that orbit circularity and velocity dispersion model results, for example, at best can only be illustrative of what the true galaxy might be like, and can not be regarded as being accurate. This deprojection issue is compounded by having to make assumptions as to what orbit initial conditions will be required to model any given galaxy. As we have seen above, these initial conditions do influence the results we are able to achieve.
\section{Summary}
\label{sec8}

Our objectives were to understand how well Schwarzschild's method is able to estimate the underlying properties of our test galaxies. We considered four properties in particular of our simulated galaxies,
\begin{enumerate}
\item the mass profile including both stellar and dark matter,
\item galaxy morphology,
\item the orbit circularity distribution of each galaxy, and
\item its velocity dispersion anisotropy.
\end{enumerate}
We have successfully met our objectives and are now in a position where we can make a recovery or estimation assessment for each property.

\begin{itemize}
\item For the total mass within one $\overline{R_{\rm e}}$, the estimate is quite good for most galaxies (eight out of nine) with absolute average relative deviations within 15 percent. The one exception is a prolate galaxy where the deviation is 36 percent. Separating the mass types, the stellar mass within one $\overline{R_{\rm e}}$ is on average $\sim24$ percent lower than the true values while the dark matter mass is $\sim38$ percent higher. These deviations are comparable with those elsewhere in the literature, notably \citet{Thomas2007}. Using a reduced galaxy sample (nine mock data sets) with a gNFW profile, these values improve to $\sim13$ percent underestimation for stellar mass and $\sim18$ percent overestimation for dark matter. Perhaps the anisotropy increase noted below is partially responsible.

\item For galaxy morphology, both the intrinsic shape parameters ($p$ and $q$) tend to be overestimated, with, with $\Delta p=0.07$ and $\Delta q=0.14$. The triaxiality parameter is generally recovered well but with model oblate and prolate galaxies tending to be slightly more triaxial than the actual test galaxies.

\item Estimates of the $\lambda_z$ circularity distribution are plausible, with the fractions of warm and counter-rotating categories being 0.05 and 0.07 overestimated, while the hot category is under-estimated by 0.12. The estimate of $\lambda_x$ is also plausible for three categories except the slowly-rotating orbits, which is on average 0.13 underestimated and dominates the bias in the hot category.

\item Comparisons of the velocity anisotropy and tangential fraction profiles with their galaxy profiles show that the galaxy models tend to be more radial in their outer regions, with the average bias in tangential fraction $f_t$ being $\sim0.064$ at one $\overline{R_{\rm e}}$ and $\sim0.1$ at higher radii. The match in the inner regions is quite good.
\end{itemize}

For the future, we are in the process of applying triaxial Schwarzschild modelling to observations of real galaxies such as in the MaNGA survey. We expect to take the findings of this paper into account in interpreting the results we achieve.
\section*{Acknowledgements}
We thank R. C. E. van den Bosch for providing us his triaxial Schwarzschild software and M. Cappellari for making his MGE software publicly available. We are also grateful to the Illustris team for the their simulated galaxy data. The modelling was accomplished on the ``Zen'' cluster at National Astronomical Observatories, Chinese Academy of Sciences (NAOC) and on ``Venus'' at Tsinghua University. This work is partly supported by the National Key Basic Research and Development Program of China (No. 2018YFA0404501 to SM), by the National Science Foundation of China (Grant No. 11821303, 11333003, 11390372 and 11761131004 to SM). LZ acknowledges support from Shanghai Astronomical Observatory, Chinese Academy of Sciences under grant NO.Y895201009. GvdV acknowledges funding from the European Research Council (ERC) under the European Union's Horizon 2020 research and innovation programme under grant agreement No. 724857 (Consolidator Grant ArcheoDyn).

\end{document}